\documentclass[preprint,12pt]{elsarticle}
\usepackage{natbib}
\usepackage{graphicx}
\usepackage{subfig}
\usepackage{pict2e}
\usepackage[T1]{fontenc}
\usepackage{fullpage}

\usepackage{amsmath}
\usepackage{amssymb}
\usepackage{theorem}
\usepackage{color}

\newcommand{\R}{{\mathcal R}}
\newcommand{\B}{{\mathcal B}}

\newcommand{\T}{{\mathcal T}}

\newcommand{\Oc}{{\mathcal O}}
\newcommand{\bt}{{\bf t}}

\newcommand{\Nb}{{\mathbb{N}}}
\newcommand{\Qb}{{\mathbb{Q}}}
\newcommand{\Rb}{{\mathbb{R}}}

\newtheorem{theorem}{Theorem}[section]
\newtheorem{lemma}[theorem]{Lemma}

\newtheorem{prop}[theorem]{Proposition}

\newtheorem{definition}[theorem]{Definition}

\theorembodyfont{\rmfamily}\newtheorem{remark}[theorem]{Remark}
\theorembodyfont{\rmfamily}\newtheorem{example}[theorem]{Example}
\DeclareMathOperator*{\argmax}{arg\,max}
\DeclareMathOperator{\blind}{Blind}

\biboptions{sort&compress,square}
\journal{Information and Computation}

\begin{document}

\begin{frontmatter}

\title{Bayesian definition of random sequences with respect to conditional probabilities\tnoteref{label1}}
 \tnotetext[label1]{Parts of the paper
were presented at the Ergod  Theory Seminar (2016 Tsukuba Univ.),
Probability Seminar (2017 Kyoto Univ.),
ISIT2017 Aachen, CCR2017 Mysour, SITA2017 Niigata,
MSJ2017 Tokyo Metropolitan Univ., MSJ2017  Yamagata Univ., 
MSJ2020 (online presentation) Nihon Univ., and MSJ2023 Chuo Univ.}
\author[rdl]{Hayato Takahashi}
\ead{hayato.takahashi@ieee.org}
\ead[url]{http://h-takahashi.sakura.ne.jp}
 \affiliation[rdl]{organization={Random Data Lab.~Inc.},
            addressline={3-8-18 Minami-Hanahata Adachi-ku},
             city={Tokyo},
             postcode={1210062},
            country={Japan}}

\begin{abstract}
We study Martin-L\"{o}f random (ML-random) points on computable probability measures on sample and parameter spaces
(Bayes models). 
We consider variants of conditional randomness defined by ML-randomness on Bayes models and those  
 of conditional blind randomness. 
We show that variants of conditional blind randomness are ill-defined 
from the Bayes statistical point of view.
We prove that if the sets of random sequences of uniformly computable parametric models are pairwise disjoint 
then there is a consistent estimator for the model.
 Finally, we present an algorithmic solution to a classical problem in Bayes statistics, i.e.~the 
posterior distributions converge weakly to almost all parameters if and only if the posterior distributions converge weakly to
all ML-random parameters.
\end{abstract}

\begin{highlights}
\item Algorithmic randomness for conditional probabilities is studied.
\item Blind randomness is ill-defined for conditional probabilities.
\item Effective orthogonality and existence of consistent estimator are equivalent.
\item An algorithmic solution to a classical problem in  Bayes statistics.
\end{highlights}


\end{frontmatter}

\section{Introduction}\label{secIntro}
We study Martin-L\"{o}f random \cite{{Kol63},{Kol65},{Kol68},{martin-lof66},{LV4th},{shenBook}}
(ML-random) points 
 on computable probability measures on sample and parameter spaces
(computable Bayes models).
We assume that samples and parameters are infinite binary sequences except for Section~\ref{secdiscuss}.
A conditional distribution \cite{{Kol33},{williams91}} is defined for a Bayes model.
A conditional distribution given a finite prefix of sample sequence is called a {\it posterior distribution}.
A marginal distribution on parameter space is called a {\it prior}.

In Bayes statistics, we study relations between samples and parameters.
Loosely speaking, we say that a probability model is a {\it true model} of a sequence or a sequence is {\it generated} by the model
if the sequence is random with respect to (w.r.t.) the model.
In Bayes statistics, we assume that a sample sequence \(x^\infty\) is generated by the marginal distribution on sample space.
Then 
we estimate the parameter \(y^\infty\) by the posterior distributions given finite prefixes of \(x^\infty\)
such that \(x^\infty\) is random w.r.t.~the conditional distribution given  \(y^\infty\).
The Bayes consistency theorem \cite{{doob48},{BreimanConiss64},{bayesNonparametrics}} says that 
the posterior distributions given finite prefixes of \(x^\infty\) weakly converge to a parameter \(y^\infty\)
for almost all \(x^\infty\) w.r.t.~the conditional distribution given \(y^\infty\)
for almost all \(y^\infty\) w.r.t.~the prior
for an appropriate class of Bayes models (consistent Bayes models).
The parameter of the true model is not determined a priori but is estimated by the posterior distributions cf.~von Mises \cite{mises}.
If we consider only a fixed parameter,
we do not know if the posterior distributions  given finite prefixes of random sequence weakly converge
to some parameter.

To state the Bayes consistency theorem for individual random sequences and parameters, we need to consider the sets of random sequences and parameters
w.r.t.~marginal distributions and the
 family of sets of random sequences w.r.t.~conditional distributions given random parameters simultaneously.
 For computable Bayes models,
we assume that (i) {\it a sequence is random w.r.t.~the marginal distribution on sample space if and only if
there is a random parameter and the sequence is random w.r.t.~the conditional distribution given the parameter}. 
For computable and consistent Bayes models, we further assume that (ii) {\it
if a sequence is random w.r.t.~the marginal distribution on sample space then
the posterior distributions given finite prefixes of the sequence weakly converge to the true model}.
The assumption (i) is equivalent to 
 that (i') {\it 
the set of random sequences w.r.t.~the marginal distribution on sample space
equals the union of the sets of random sequences w.r.t.~the conditional distributions given random parameters.}
The family of the sets of random sequences w.r.t.~the conditional distributions given random parameters is determined by 
those distributions (and hence by the Bayes model).
The assumption (i') requires that
the union of those sets
is determined only by the marginal distribution on sample space.
Assumption (i), (i'), and (ii) are natural requirements for any notion of randomness in Bayes models,
see Section~\ref{discussBayesDef}.

Theorem~\ref{conditional-prob}  \cite{{takahashiISIT2006},{takahashiRIMS2008},{takahashiIandC}} shows that for a computable Bayes model
there exists a version of  conditional distribution (the standard conditional distribution)
that is defined for ML-random parameters w.r.t.~the prior.
We study  four families of the  sets of random sequences w.r.t.~the standard conditional distribution:
two of them are defined by ML-randomness on Bayes models (variants of conditional Bayes randomness) and the others 
are defined by the blind (Hippocratic) tests \cite{{Bienvenu2011},{kjoshanssen}} for conditional 
distributions with given parameters (variants of conditional blind randomness).
Theorem~\ref{thMixcond} parts 1 and 3 show that 
variants of conditional Bayes randomness satisfy the assumptions (i') and (ii) for all computable Bayes models.
On the other hand, there are different computable and consistent Bayes models that
have the same marginal distribution on sample space but 
 their unions of conditional blind random sequences for all random parameters are different.
Theorem~\ref{thMixcond} part 4 shows that
there is no family of sets of random sequences w.r.t.~the marginal distribution on sample space
that satisfy the assumption (i') for variants of conditional blind randomness for the class of computable and
consistent Bayes models. 
We consider that variants of conditional blind randomness are ill-defined as conditional randomness for Bayes models.

The rest of the paper is structured as follows.
In Section~\ref{secDefCondML}, we state our main theorem.
In Section~\ref{sec2}, we study ML-random points in Bayes models.
In Section~\ref{subsecBayesrandom}, we 
show the relationship between variants of random sequences w.r.t.~the standard conditional distributions.
In this paper, we assume the computability of joint probability measures, however, we  
do not demand uniform computability on standard conditional distributions.
In Section~\ref{sec-ML-consis}, we compare ML-randomness in  Bayes models 
with randomness for uniformly computable parametric models \cite{{levinUniform},{Bienvenu2011},{gacs2005},{kjoshanssen},{vovkandvyugin93}}.
Theorem~\ref{propBayesUniform} shows that standard conditional distributions are equal to uniformly computable parametric models on ML-random parameters
when the Bayes models are constructed from uniformly computable parametric models and computable priors.
We show that if the sets of uniform random sequences are pairwise disjoint for different parameters
(effectively orthogonal \cite{Bienvenu2011}), then there is a consistent estimator, i.e.~a measurable function
that equals \(y^\infty\) with probability one for the probability model of the parameter  \(y^\infty\) for all \(y^\infty\).
In Section~\ref{secdiscuss}, we discuss ML-randomness of Bayes models on complete separable metric spaces and
present an algorithmic solution to a classical problem in Bayes statistics \cite{diaconisFreedman1986}, 
i.e.~for the Bayes models on complete separable metric spaces, 
the posterior distribution is consistent at almost all parameters if and only if the posterior distribution is consistent
at all ML-random parameters. Finally in Section~\ref{discussBayesDef}, 
we briefly re-discuss randomness in statistical models.
We compare our notion of randomness in Bayes models with that in non-Bayes models (parametric models without prior),
which may help to understand our assumptions (i), (i'), and (ii).

\section{Main theorem}\label{secDefCondML}
Let  \(\Omega\) be the set of infinite binary sequences, and \(S\)  the set of finite binary strings.
We also call an element of \(\Omega\cup\Omega^2\) a point. 
Let  \(\Delta(x)\) be the set of infinite binary sequences that start with \(x\in S\), and \(|x|\) the length of \(x\in S\).
To clarify the difference between finite strings and infinite sequences, we use symbols such as  \(x, y\) for finite strings and  \(x^\infty, y^\infty\) for infinite sequences.
Except for the clearly stated cases, 
do not confuse symbols such as  \(x^\infty\) with the repetition of a string.
Let \(\lambda\) be the empty word.  
We write \(x\sqsubseteq y\)  if \(x\) is a prefix of \(y\) for \(x,y\in S\cup\Omega\) including the case \(x=y\) and
 \(x\sqsubset y\) if \(x\sqsubseteq y\) and \(x\ne y\).
 Let \(B^c\) be the complement of  \(B\).
 We write \(\tilde{A}:= \bigcup_{x\in A}\Delta(x)\) for \(A\subseteq S\).
For \(A\subseteq S^2\),
\(\tilde{A}\)  is defined similarly. 

Let open sets in \(\Omega\) be those generated by \(\{\Delta(x) \mid x\in S\}\), 
i.e.~every open set  in \(\Omega\) is a union of elements in \(\{\Delta(x) \mid x\in S\}\).
Let \((\Omega,\B_1)\) be a measurable space where \(\B_1\) is the smallest \(\sigma\)-algebra that includes 
\(\{\Delta(x) \mid x\in S\}\).
Similarly, let open sets in \(\Omega^2\) be those generated by  \(\{\Delta(x)\times\Delta(y) \mid x, y\in S\}\) and
 \((\Omega^2,\B_2)\)  a measurable space where \(\B_2\) is the smallest  \(\sigma\)-algebra that includes
\(\{\Delta(x)\times\Delta(y)\mid x,y\in S\}\).
Let 
\(d(x^\infty,y^\infty):= \sum_n \vert x_n-y_n\vert 2^{-n}\) for all \(x^\infty,y^\infty\in\Omega\) where 
\(x^\infty=x_1 x_2\cdots\) and
\(y^\infty=y_1y_2\cdots\).
Then \(d\) is a metric on \(\Omega\) and open sets induced by \(d\) are equivalent to those generated by 
\(\{\Delta(x) \mid x\in S\}\). 
Similarly for \((x^\infty,y^\infty) ,(x^{\prime\infty},y^{\prime\infty})\in\Omega^2\),  
let \(d^2((x^\infty,y^\infty) ,(x^{\prime\infty},y^{\prime\infty}) ):= d(x^\infty,x^{\prime \infty})+d(y^\infty,y^{\prime\infty})\).
Then \(d^2\) is a metric on \(\Omega^2\) and open sets induced by \(d^2\) are equivalent to those generated by 
\(\{\Delta(x)\times\Delta(y)\mid x,y\in S\}\).
In the following, we consider the metrics \(d\)  on \(\Omega\) and \(d^2\) on \(\Omega^2\).

Except for Section~\ref{secdiscuss}, 
set \(X=Y=\Omega\).
Let \(P\) be a probability measure on  \((X\times Y, \B_2)\), \(P_X\) marginal distribution on 
\((X,\B_1)\), and \(P_Y\) marginal distribution on  \((Y,\B_1)\).
For all \(x,y\in S\), let  \(P(x,y):= P(\Delta(x)\times\Delta(y))\), \(P_X(x):= P(\Delta(x)\times \Omega)\), and \(P_Y(y):= P(\Omega \times \Delta(y))\).
For all \(x,y\in S\), let  \(P(x\mid y):= P(x,y)/P_Y(y)\) if \(P_Y(y)\ne 0\) and \(P_{Y|X}(y\mid x):= P(x,y)/P_X(x)\) (posterior distribution)   if \(P_X(x)\ne 0\).

Let \(U\subseteq S\times \Nb\).  
 \(U\) is called a {\it test} (effective null set) or ML-test w.r.t.~\(P\) on \((\Omega, \B)\) if 
 \(U\) is recursively enumerable (r.e.),
 \(\ \tilde{U}_n\supseteq \tilde{U}_{n+1}\), and \(P(\tilde{U}_n)< 2^{-n}\),
 where \(U_n=\{ x \mid (x, n)\in U\}\), for all \(n\).
The  ML-random sequences w.r.t.~\(P\) are defined as the complement of the effective null sets w.r.t.~\(P\). We denote it by \(\R^P\), i.e.~\(\R^P:=  (\bigcup_{U\colon test}\bigcap_n \tilde{U}_n)^c\).
 Let 
\(\R^{P, A}:=  (\bigcup_{U\colon test}\bigcap_n \tilde{U}_n)^c\), where \(U\) is a test with oracle \(A\), 
i.e.~\(U\) is r.e.~with oracle \(A\),
 \( \tilde{U}_n\supseteq \tilde{U}_{n+1}\), and \(P(\tilde{U}_n)< 2^{-n}\) for all \(n\).
 Similarly,  \(\R^P\) and \(\R^{P,A}\) are defined w.r.t.~\(P\) on \((\Omega^2, \B_2)\).
\(\R^P\) and \(\R^{P,A}\) are forms of blind randomness, i.e.~we neither assume that \(P\) is computable nor
 computable with oracle \(A\).
In the following, for simplicity, we say that 
\((\Omega,\B_1,P)\) and \((X\times Y,\B_2,P)\) are  computable if \(P\) on \((\Omega,\B_1)\) and \((\Omega^2,\B_2)\) are computable, respectively.

We obtain the following theorem from 
the martingale convergence theorem for individual ML-random sequences  \cite{{takahashiISIT2006},{takahashiIandC}}.
\begin{theorem}[Takahashi \cite{{takahashiISIT2006},{takahashiRIMS2008},{takahashiIandC}}]\label{conditional-prob}
Assume that \((X\times Y, \B_2,P)\) is computable. 
For all \(x\in S\) and \(y^\infty\in\R^{P_Y}\), set
\[
P(x\mid y^\infty):= \lim_{y\to y^\infty}P(x\mid y)
\]
if the right-hand side exists. Then for each \(y^\infty\in\R^{P_Y}\),
\(P(\cdot\mid y^\infty)\) is a probability measure on \((X, \B_1)\).
\end{theorem}
\begin{definition}\label{standardDef}
The family of probability measures \(\{P(\cdot\mid y^\infty)\mid y^\infty\in\R^{P_Y}\}\) is called the
 {\it standard conditional distribution}.
\end{definition}
For each  \(x\in S\), conditional probability
 \(P(x\mid y^\infty)\) is a Borel-measurable random variable  on \(Y\) that satisfies
\[\int_{\Delta(y)}P(x\mid y^\infty)dP_Y=P(x\mid y)\text{ for all }y\in S,\]
 see Theorem 33.1 on page pp.434 in \cite{pbmBill}.
Conditional probability is not unique. 
Two versions of conditional probability are called equivalent if they coincide for almost all \(y^\infty\).
From the Radon-Nikod\'{y}m theorem, the standard conditional distribution is a version of conditional probability \cite{williams91}.
A version of conditional probability is called {\it regular} if the conditional probability given the parameter is a probability measure 
for almost all parameters.
By Theorem~\ref{conditional-prob},  the standard conditional distribution  is  
 a probability measure for each ML-random parameter and hence regular, but
it may not be computable with oracle access to the ML-random parameter, see Remark~\ref{remarkUnifrom}.
To state the theorems in the Bayes models for individual ML-random points,
 e.g.~consistency theorem for the posterior  distributions (Theorem~\ref{th-consis-pos}),
we fix the standard conditional distribution as a version of conditional probability.
A computable Bayes model defines a standard conditional distribution and prior, and vice versa. 

We define consistent Bayes models.
\begin{definition}
Let  \(P\) be a probability measure on \((X\times Y,\B_2)\).
Let  \(\delta_{y^\infty}\) be  the probability measure on \((Y,\B_1)\) such that  
\(\delta_{y^\infty}(\{y^\infty\})=1\). 
The posterior distribution \(P_{Y|X}(\cdot\mid x)\) weakly converges  to \(\delta_{y^\infty}\) as \(x\to x^\infty\) 
 if and only if for any neighborhood \(A\) of \(y^\infty\), \(\lim_{x\to x^\infty}P_{Y|X}(A\mid x)=1\).
The posterior distribution is called {\it consistent at \(y^\infty\)} if it weakly converges to \(\delta_{y^\infty}\) as \(x\to x^\infty\) 
for almost all \(x^\infty\) w.r.t.~\(P(\cdot\mid y^\infty)\).
\(P\) is called consistent if the posterior distribution is consistent at almost all parameters.
\end{definition}

For a set \(A\subseteq \Omega\times \Omega\), let \(A_{y^\infty}:= \{ x^\infty\mid (x^\infty,y^\infty)\in A\}\).
Consider the following   sets  for  \(y^\infty\in\R^{P_Y}\),\\
(i) \(\bigcap_{y\to y^\infty}\R^{P(\cdot\mid y)}\), the intersection of the sets of ML-random sequences w.r.t.~\(P(\cdot\mid y)\) for all finite prefixes of \(y^\infty\),\\
(ii)  \(\R^P_{y^\infty}\), the section of ML-random points  \(\R^P\) at  \(y^\infty\),\\
(iii) \(\R^{P(\cdot\mid y^\infty)}\), ML-random sequences w.r.t.~\(P(\cdot\mid y^\infty)\), and\\
(iv) \(\R^{P(\cdot\mid y^\infty),y^\infty}\), ML-random sequences w.r.t.~\(P(\cdot\mid y^\infty)\) with oracle \(y^\infty\).\\
Figure~\ref{figA} shows relations between these sets.

\begin{figure}
\centering
\subfloat[General Case]{
\begin{picture}(100,140)(20,0)
\put(40, 130){\(\bigcap_{y\to y^\infty} \R^{P(\cdot\mid y)}\)}
\put(35, 100){\rotatebox{50}{\(\subseteq\)}}
\put(10, 100){{\footnotesize (i)}}
\put(70, 100){\rotatebox{130}{\(\subseteq\)}}
\put(95, 100){{\footnotesize (ii)}}
\put(20,  70){\(\R^P_{y^\infty}\)}
\put(80,70){\(\R^{P(\cdot\mid y^\infty)}\)}
\put(35, 40){\rotatebox{130}{\(\subseteq\)}}
\put(10, 40){{\footnotesize (iii)}}
\put(70, 40){\rotatebox{50}{\(\subseteq\)}}
\put(95, 40){{\footnotesize (iv)}}
\put(40, 10){\(\R^{P(\cdot\mid y^\infty),y^\infty}\)}
\end{picture}
}
\subfloat[Consistent Case]{
\begin{picture}(100,140)(20,0)
\put(45,  130){\(\R^P_{y^\infty}=\bigcap_{y\to y^\infty} \R^{P(\cdot\mid y)}\)}
\put(50, 100){\rotatebox{90}{\(\subseteq\)}}
\put(75, 100){{\footnotesize (v)}}
\put(45,70){\(\R^{P(\cdot\mid y^\infty)}\)}
\put(50, 40){\rotatebox{90}{\(\subseteq\)}}
\put(75, 40){\footnotesize (vi)}
\put(45, 10){\(\R^{P(\cdot\mid y^\infty),y^\infty}\)}
\end{picture}
}
\caption{
Figure (a) shows relations between variants of conditional random sequences.  Figure (b) shows
those relations when the Bayes model is consistent. 
If the conditional probability  is computable with  ML-random oracle \(y^\infty\) and 
 the Bayes model is consistent then the four sets in figure (b) are equal. 
Theorem~\ref{counter-example} shows a counter-example of the equalities in (ii) and (v).
Lemma~\ref{lemEX} shows a counter-example of the equalities in (i) and (iv).
 Lemma~\ref{lemEX} and Theorem~\ref{counter-example} show that \(\R^P_{y^\infty}\) and \(\R^{P(\cdot\mid y^\infty)}\) are incomparable in the general case.
The counter-example of equality in (iii) is due to Bauwens \cite{{BST2016},{Bauwens2015}}.
The equality in (vi) for  consistent parametric models was presented in Kjos Hanssen \cite{kjoshanssen} and  Bienvenu et al.~\cite{Bienvenu2011}, see Section~\ref{sec-ML-consis}.
An example of the equality in (vi) for non-uniformly computable orthogonal conditional distributions is presented in Theorem~\ref{counter-example}.
It remains open whether the equality in (vi) always holds for all  ML-random \(y^\infty\) (Remark~\ref{remarkopen}).
}\label{figA}
\end{figure}

\begin{theorem}[Main theorem]\label{thMixcond} \hfill\\
1.~Assume that   \((X\times Y, \B_2, P)\) is computable. Then
\begin{align*}
\R^{P_X}&=\bigcup_{y^\infty\in\R^{P_Y}}\bigcap_{y\to y^\infty}\R^{P(\cdot\mid y)}=
\bigcup_{y^\infty\in\R^{P_Y}} \R^P_{y^\infty}
\supseteq \bigcup_{y^\infty\in\R^{P_Y}}\R^{P(\cdot\mid y^\infty)}\supseteq
\bigcup_{y^\infty\in\R^{P_Y}}\R^{P(\cdot\mid y^\infty), y^\infty}.
\end{align*}
2.~Assume that   \((X\times Y, \B_2, P)\) is computable and
 \(P(\cdot\mid y^\infty)\) is computable with oracle \(y^\infty\) for each fixed \(y^\infty\in\R^{P_Y}\).
Then
\[
\R^{P_X}=\bigcup_{y^\infty\in\R^{P_Y}}\bigcap_{y\to y^\infty}\R^{P(\cdot\mid y)}=
\bigcup_{y^\infty\in\R^{P_Y}} \R^P_{y^\infty}
= \bigcup_{y^\infty\in\R^{P_Y}}\R^{P(\cdot\mid y^\infty)}=
\bigcup_{y^\infty\in\R^{P_Y}}\R^{P(\cdot\mid y^\infty), y^\infty}.
\]
3.~Assume that \((X\times Y,\B_2,P)\)  is computable and consistent.
Then 
\[\begin{split}
 x^\infty\in\R^{P_X}\text{ if and only if }&\text{there is a unique }y^\infty\in\R^{P_Y}\text{ such that}\\
&  P_{Y|X}(x)\text{ weakly converges to }\delta_{y^\infty}\text{ as }x\to x^\infty \text{and}\\
& x^\infty\in\R^P_{y^\infty} (=\bigcap_{y\to y^\infty}\R^{P(\cdot\mid y)}).
\end{split}
\]
4.~There is
no \(\tilde{\R}\colon \{P\mid (\Omega,\B_1,P)\text{ is computable}\}\to\{A\mid A\subseteq\Omega\}\)
such that
\[\tilde{\R}(P_X)=\bigcup_{y^\infty\in\R^{P_Y}}\R^{P(\cdot\mid y^\infty)}\text{ for all computable and consistent }(X\times Y,\B_2, P).\]
Similarly, there is no \(\tilde{\R}\colon \{P\mid (\Omega,\B_1,P)\text{ is computable}\}\to\{A\mid A\subseteq\Omega\}\)
such that 
\[\tilde{\R}(P_X)=\bigcup_{y^\infty\in\R^{P_Y}}\R^{P(\cdot\mid y^\infty),y^\infty}\text{ for all computable and consistent }(X\times Y,\B_2, P).\]
5.~Assume that   \((X\times Y, \B_2, P)\) is computable. Then\\
the sets (i)--(iv) are Borel sets in \(\Omega\) and 
the probabilities of these sets w.r.t.~\(P(\cdot\mid y^\infty)\) are one
for all \(y^\infty\in\R^{P_Y}\), and\\
\(\bigcup_{y^\infty\in\R^{P_Y}}\bigcap_{y\to y^\infty}\R^{P(\cdot\mid y)}\) and 
\(\bigcup_{y^\infty\in\R^{P_Y}} \R^P_{y^\infty}\) are Borel sets in \(\Omega\) and
the probabilities of these sets w.r.t.~\(P_X\) are one.
\end{theorem}
The proof of Theorem~\ref{thMixcond} is given after the proof of Theorem~\ref{counter-example}.

Roughly speaking, the sets (i) and (ii) are defined from Bayes models; those
(iii) and (iv) are defined from conditional probabilities with given parameters.
For simplicity, we call the families of sets \(\{ \bigcap_{y\to y^\infty}\R^{P(\cdot\mid y)} \mid y^\infty\in\R^{P_Y}\}\) and 
\(\{\R^P_{y^\infty}\mid y^\infty\in\R^{P_Y}\}\) {\it variants of conditional Bayes randomness}.
We call the families of sets 
\(\{\R^{P(\cdot\mid y^\infty)} \mid y^\infty\in\R^{P_Y}\}\) and 
\(\{\R^{P(\cdot\mid y^\infty),y^\infty} \mid y^\infty\in\R^{P_Y}\}\)
{\it variants of conditional blind randomness}.
By Theorem~\ref{thMixcond} parts 1 and 3,
both of the variants of conditional Bayes randomness
satisfy our assumptions (i') and (ii).
On the other hand, 
by Theorem~\ref{thMixcond} part 4, 
neither variant of conditional blind randomness satisfies the assumption (i').
We consider that variants of conditional blind randomness are ill-defined as conditional randomness for Bayes models. 

Takahashi \cite{{takahashiISIT2006},{takahashiRIMS2008},{takahashiIandC},{takahashiIandC2}} defined
the family of sets \(\{\R^P_{y^\infty}\mid y^\infty\in\R^{P_Y}\}\) to be the sets of 
conditional random sequences 
w.r.t.~the standard conditional distribution \(\{P(\cdot\mid y^\infty)\mid y^\infty\in\R^{P_Y}\}\).
\begin{remark}\label{measurabilityremark}
The author does not know if  the sets 
\(\bigcup_{y^\infty\in\R^{P_Y}}\R^{P(\cdot\mid y^\infty)}\) and \(\bigcup_{y^\infty\in\R^{P_Y}}\R^{P(\cdot\mid y^\infty),y^\infty}\)
 are always measurable for all computable \((X\times Y, \B_2, P)\).
\end{remark}
\section{ML-randomness and Bayes models}\label{sec2}
The van Lambalgen theorem \cite{lambalgen87} states that a pair of sequences \((x^\infty,y^\infty)\in\Omega^2\) is ML-random w.r.t.~the product  of uniform measures if and only if  \(y^\infty\) is ML-random and \(x^\infty\) is ML-random with oracle  \(y^\infty\).
\begin{theorem}[Generalized van Lambalgen theorem, Takahashi \cite{{takahashiISIT2006},{takahashiRIMS2008},{takahashiIandC},{takahashiIandC2}}]\label{col-ifpart}
Assume that  \((X\times Y, \B_2,P)\) is computable. Then
\begin{equation}\label{genLambalA}
\R^P_{y^\infty}\supseteq\R^{P(\cdot\mid y^\infty),y^\infty} \text{ for all }y^\infty\in\R^{P_Y}.
\end{equation}
Fix \(y^\infty\in\R^{P_Y}\) and assume that \(P(\cdot\mid y^\infty)\) is computable with oracle \(y^\infty\).
Then
\begin{equation}\label{generalLambal}
\R^P_{y^\infty}=\R^{P(\cdot\mid y^\infty),y^\infty}.
\end{equation}
\end{theorem}
\begin{theorem}[Takahashi \cite{{takahashiISIT2006},{takahashiRIMS2008},{takahashiIandC}}]\label{bayes-mix}
Assume that \((X\times Y,\B_2,P)\) is computable. Then
\[
P(\R^P_{y^\infty}\mid y^\infty)=1\text{ if }y^\infty\in\R^{P_Y}\text{ and }\R^P_{y^\infty}=\emptyset\text{ else.}
\]
\begin{equation}\label{eqMLMarginal}
\R^{P_X}=\bigcup_{y^\infty\in\R^{P_Y}}\R^P_{y^\infty}.
\end{equation}
\end{theorem}
\begin{remark}\label{remarkUnifrom}
1.~Vovk and V'yugin \cite{vovkandvyugin93} proved     (\ref{generalLambal}) 
and (\ref{eqMLMarginal}) for Bayes models that are constructed from uniformly computable parametric models and computable priors, see Section~\ref{sec-ML-consis}.\\
\noindent
2.~Theorem 5.2 in \cite{takahashiIandC}  demonstrates   equation  (\ref{genLambalA})  when \(P(\cdot\mid y^\infty)\) is  computable with oracle \(y^\infty\).
However,  the same proof for    (\ref{genLambalA}) holds true when \(P(\cdot\mid y^\infty)\) is not  computable with oracle \(y^\infty\).
In \cite{takahashiIandC2}, (\ref{generalLambal}) is proved without assuming uniform computability of the conditional distributions. \\
\noindent 
3.~Non-computable conditional distributions are presented in work by Ackerman et al.~\cite{roy2011confpaper}.
 Bauwens  \cite{Bauwens2015}  showed an example that violates the equality in (\ref{generalLambal})
 and  \(\R^{P(\cdot\mid y^\infty),y^\infty}\) is a proper subset of \(\R^P_{y^\infty}\) in (\ref{genLambalA})
when the conditional distribution is not computable with oracle \(y^\infty\).
Takahashi \cite{takahashi2014} showed an example that the  conditional distributions are not computable with oracle \(y^\infty\)
 for all \(y^\infty\),  but    (\ref{generalLambal}) holds true.
For more details on the generalized van Lambalgen theorem, see the survey Bauwens et al.~\cite{BST2016}.
\qed
\end{remark}

Two probability measures \(P\) and \(Q\) on \((\Omega,\B_1)\)  are called  {\it orthogonal} and denoted by \(P\perp Q\) 
if there is \(A\in\B_1\) such that \(P(A)=1\) and \(Q(A)=0\).
The following theorem shows equivalent statements for the consistency of Bayes models including statements described with ML-randomness. 
\begin{theorem}\label{th-consis-pos}
Assume that  \((X\times Y,\B_2,P)\) is computable. 
The following  statements are equivalent:\\
\textnormal{(i)} \( P(\cdot\mid  y)\perp  P(\cdot\mid  z)\) 
for all  \(\Delta( y)\cap\Delta( z)=\emptyset\).\\
\textnormal{(ii)} \(\R^{ P(\cdot\mid  y)}\cap\R^{ P(\cdot\mid  z)}=\emptyset\)
for all \(\Delta( y)\cap\Delta( z)=\emptyset\).\\
\textnormal{(iii)} 
\(P_{Y\vert X}(\cdot\mid x)\) weakly converges  to \(\delta_{y^\infty}\) as \(x\to x^\infty\) for all \((x^\infty,y^\infty)\in\R^P\).\\
\textnormal{(iv)} \(\R^{P_Y}\subseteq\{y^\infty\in\Omega \mid \text{the posterior distribution }P_{Y|X}\text{ is consistent at }y^\infty\}\).\\
\textnormal{(v)} \(P\) is consistent.\\
\textnormal{(vi)} There is a measurable \(f\colon X\to Y\) such that 
\(P(f^{-1}(y^\infty)\mid y^\infty)=1\) for almost all \(y^\infty\) w.r.t.~\(P_Y\).\\
\textnormal{(vii)} \(\R^{ P}_{ y^\infty}\cap\R^{ P}_ {z^\infty}=\emptyset\) for all \(y^\infty\ne z^\infty\).\\
\textnormal{(viii)} There is a measurable onto \(f\colon \R^{P_X}\to \R^{P_Y}\)  such that 
\(P(f^{-1}(y^\infty)\mid y^\infty)=1\) for  all \(y^\infty\in\R^{P_Y}\).
\end{theorem}
Before we prove the theorem, we show a lemma.
\begin{lemma}\label{bayesconsislemmaA}
Assume that \((X\times Y,\B_2,P)\) is consistent.
Then there is a measurable \(f\colon X\to Y\) such that 
\(P(f^{-1}(y^\infty)\mid y^\infty)=1\) for almost all \(y^\infty\) w.r.t.~\(P_Y\).
\end{lemma}
Proof)
Let 
\begin{align*}
g^{-1}(y^\infty)&:= \{x^\infty\mid P_{Y|X}(\cdot\mid x)\text{ weakly converges to }\delta_{y^\infty}\text{ as }x\to x^\infty\}
\text{ and }\\
f^{-1}(y^\infty)&:= \left\{
\begin{array}{cl}
g^{-1}(y^\infty) & \text{if }y^\infty\ne 0^\infty\\
g^{-1}(0^\infty)\cup (X\setminus \bigcup_{y^\infty\in Y}g^{-1}(y^\infty)) & \text{else},
\end{array}\right.
\end{align*}
where \(0^\infty\) is the sequence consisting of 0s.

Then \(f\colon X\to Y\) is well-defined, i.e.~\(f^{-1}(y^\infty)\cap f^{-1}(z^\infty)=\emptyset\) for all \(y^\infty\ne z^\infty\).
We show that \(f\) is measurable. 
Let 
\begin{align*}
A_{y,\epsilon,L}&:= \{x\mid P_{Y|X}(y\mid x)>1-\epsilon,\ |x|=L\}\text{ and }\\
B_y&:= \bigcap_{k\colon \epsilon=1/k}\bigcup_L\bigcap_{L^\prime\colon L\leq L^\prime}\tilde{A}_{y,\epsilon,L^\prime}.
\end{align*}
We have that 
\(B_y=\{x^\infty\mid \lim_{x\to x^\infty}P_{Y|X}(y\mid x)=1\}\),
\(B_y\cap B_w=\emptyset\) if \(\Delta(y)\cap\Delta(w)=\emptyset\), and 
\(B_y\supseteq B_w\text{ if }y\sqsubseteq w\).
By the definition of weak convergence of the posterior distributions, we have
\[\bigcap_{n\colon |z|\leq n}\bigcup_{\substack{y\colon |y|=n\\ z\sqsubseteq y}} B_y=g^{-1}(\Delta(z)).\]
Since \(\tilde{A}_{y,\epsilon,L}\) is open for all \(y,\epsilon, L\), we have that
 \(g^{-1}(\Delta(z))\in\B_1\) for all \(z\in S\) and \(f\) is measurable.
Since \(P\) is consistent, we have
\(P(f^{-1}(y^\infty)\mid y^\infty)=1\) for almost all \(y^\infty\).
\qed

\noindent
Proof of Theorem~\ref{th-consis-pos})
For the proof of (i) \(\Rightarrow\) (ii) \(\Rightarrow\) (iii), see the proof of Theorem 6.1 in \cite{takahashiIandC2}.
The implication (iii) \(\Rightarrow\) (iv) is immediate.
The implication (iv) \(\Rightarrow\) (v) follows from that \(P_Y(\R^{P_Y})=1\).
The implication (v) \(\Rightarrow\) (vi) is due to Lemma~\ref{bayesconsislemmaA}.
We show (vi) \(\Rightarrow\) (i).
By (vi), we have \(f^{-1}(\Delta(y)), f^{-1}(\Delta(z))\in\B_1\) and \(f^{-1}(\Delta(y))\cap f^{-1}(\Delta(z))=\emptyset\)
for all \(\Delta(y)\cap\Delta(z)=\emptyset\).
By the Fubini theorem and (vi), we have 
\(P(f^{-1}(\Delta(y))\mid y)=1\) and \(P(f^{-1}(\Delta(z))\mid z)=1\) for all \(\Delta(y)\cap\Delta(z)=\emptyset\).
We have the equivalence (i)--(vi).

The implication (iii) \(\Rightarrow\) (vii) is immediate. We show (vii) \(\Rightarrow\) (viii).
Let \(f^{-1}(y^\infty)=\R^P_{y^\infty}\) for all \(y^\infty\in\R^{P_Y}\).
By (vii) and Theorem~\ref{bayes-mix}, \(f\colon \R^{P_X}\to\R^{P_Y}\) is well-defined, an onto function, and
\(P(f^{-1}(y^\infty)\mid y^\infty)=1\) for  all \(y^\infty\in\R^{P_Y}\).
By Theorem~\ref{bayes-mix}, 
\(f^{-1}(\R^{P_Y})=\R^{P_X}\in\B_1\) and similarly, \(f^{-1}(\Delta(s)\cap \R^{P_Y})\in\B_1\) for any \(s\in S\), and 
we see that \(f\) in (viii) is measurable. 
The proof of (viii) \(\Rightarrow\) (i) is similar to that of (vi) \(\Rightarrow\) (i).
\qed
\begin{remark}
 Takahashi \cite{{takahashiISIT2006},{takahashiRIMS2008},{takahashiIandC2}} showed the equivalence of the statement in 
Theorem~\ref{th-consis-pos}
except for statements (iv) and (v).
The equivalence   (i) and (vi) is due to Corollary 2 in Breiman et al.~\cite{BreimanConiss64}.
The equivalence  (i) and (ii) is due to Martin-L\"{o}f (pp.~103 second paragraph \cite{martin-lof68}),
see also Theorem 4.1 in \cite{takahashiIandC2} and Theorem 8.6 in \cite{Bienvenu2011}.
Doob  \cite{doob48} showed that if there is a measurable function \(f\colon X\to Y\) such that 
\(P(f^{-1}(y^\infty)\mid y^\infty)=1\) for  all \(y^\infty\in Y\) then \(P\) is consistent, see also 
\cite{{bayesNonparametrics},{schwartz65}}.
\end{remark}
For further study on ML-random points on Bayes models, see  D\k{e}bowski~\cite{debowski}, Takahashi~\cite{{takahashiIandC},{takahashiIandC2}} and Vovk and
V'yugin~\cite{vovkandvyugin93}.
When the prior is discrete, Theorem~\ref{th-consis-pos} reduces to the consistency of the MDL model 
selection~\cite{{barronphd},{barron-et-al}}
 for individual random sequences~\cite{{takahashiIandC2}}.
 Li and Vit\'{a}nyi \cite{LV4th} study MDL model selection in terms of Kolmogorov complexity.
\subsection{Relations between conditional randomness}\label{subsecBayesrandom}
We prove all inclusions shown in Figure \ref{figA}.
\begin{lemma}\label{lemma-A}
Assume that \((X\times Y,\B_2,P)\) is computable. 
Then
\[\R^{P(\cdot\mid y^\infty)}\subseteq\R^{P(\cdot\mid y)}
\text{ for all }y\sqsubset y^\infty\text{ and }y^\infty\in\R^{P_Y}.\]
\end{lemma}
Proof)
First we prove the lemma for \(y=\lambda\).
Let \(y=\lambda\) then \(P(\cdot\mid y)=P_X\).
Let \(U^X\) be a test w.r.t.~\(P_X\) and  \(U^{X\times \lambda}:= \{(x,\lambda, n)\mid (x, n)\in U^X\}\).
\(U^{X\times \lambda}\) is a test w.r.t.~\(P\).
Since \((\tilde{U}^{X\times \lambda}_n)_{y^\infty}=\tilde{U}^X_n\), 
from  Corollary~4.1 in \cite{takahashiIandC},
for all \(y^\infty\in\R^{P_Y}\) 
there is an integer \(M\) such that 
\[P(\sum_n I_{\tilde{U}^X_n}> k\mid y^\infty) < \frac{M}{k}\text{ for all }k,\]
where \(I\) is the characteristic function, i.e.~\(I_{\tilde{U}_n^X}(x^\infty)=1\) if \(x^\infty\in \tilde{U}^X_n\) and  \(0\) otherwise.
We have a test \(V^X\) (\(\tilde{V}^X_k:=\{\sum_n I_{\tilde{U}^X_n} > M2^k\}\text{ for all }k\)) w.r.t.~\(P(\cdot\mid y^\infty)\)  and the lemma is proved for \(y=\lambda\).
Similarly, we can show the lemma for all finite prefix \(y\sqsubset y^\infty\).
\qed
\begin{theorem}\label{mutual-singular}
Assume that  \((X\times Y,\B_2,P)\) is computable. 
\begin{equation}\label{subSet}
\bigcap_{y\to y^\infty}\R^{P(\cdot\mid y)}\supseteq \R^{P(\cdot\mid y^\infty)}\text{ and }  \bigcap_{y\to y^\infty}\R^{P(\cdot\mid y)}\supseteq \R^P_{y^\infty} \text{ for all }y^\infty\in\R^{P_Y}.
\end{equation}
If  \(P\) is consistent, we have
\begin{equation}\label{eq-singular-case}
\bigcap_{y\to y^\infty}\R^{P(\cdot\mid y)}=\R^P_{y^\infty}\supseteq\R^{P(\cdot\mid y^\infty)}\supseteq\R^{P(\cdot\mid y^\infty),y^\infty} \text{ for all }y^\infty\in\R^{P_Y}.
\end{equation}
Fix \(y^\infty\in\R^{P_Y}\) and assume that \(P(\cdot\mid y^\infty)\) is computable with oracle \(y^\infty\).
If \(P\) is consistent, we have
\begin{equation}\label{singularEquality}
\bigcap_{y\to y^\infty}\R^{P(\cdot\mid y)}=\R^P_{y^\infty}=\R^{P(\cdot\mid y^\infty)} =\R^{P(\cdot\mid y^\infty),y^\infty}.
\end{equation}
\end{theorem}
\noindent
Proof)
Similarly,  as presented  in Theorem~\ref{bayes-mix}, we can show that 
\begin{equation}\label{bayes-mixtureB}
\R^{P(\cdot\mid y)}=\bigcup_{y^\infty\in\R^{P_Y}\cap\Delta(y)}\R^P_{y^\infty}\text{ for all }y.
\end{equation}
From Lemma~\ref{lemma-A}, we obtain   (\ref{subSet}).

Assume that \(P\) is consistent.
By Theorem~\ref{th-consis-pos}, we have \(\R^{ P}_{ y^\infty}\cap\R^{ P}_ {z^\infty}=\emptyset\) if \( y^\infty\ne z^\infty\).
From   (\ref{bayes-mixtureB}), we have
\begin{equation}\label{singularSubEq}
\bigcap_{y\to y^\infty} \R^{P(\cdot\mid y)}= \R^P_{y^\infty}.
\end{equation}
From  (\ref{subSet}) and (\ref{singularSubEq})  we have   (\ref{eq-singular-case}).
Eq.~(\ref{singularEquality}) follows from Theorem~\ref{col-ifpart} and  (\ref{eq-singular-case}).
\qed

The following lemma shows a counterexample that the inclusions  (i) and (iv) in Figure \ref{figA} can be strict simultaneously.
\begin{lemma}\label{lemEX}
Let \(P:= U\times U\), where \(U\) is the uniform probability measure on \((\Omega,\B_1)\), then
\(\bigcap_{y\to y^\infty}\R^{P(\cdot\mid y)}\setminus\R^P_{y^\infty}\ne\emptyset\) and 
\(\R^{P(\cdot\mid y^\infty)}\setminus\R^{P(\cdot\mid y^\infty),y^\infty}\ne\emptyset\)
for all \(y^\infty\in\R^U\).
\end{lemma}
Proof)
The diagonal set is covered by a test, i.e.~\(\{(x^\infty, x^\infty)\mid x^\infty\in\Omega\}\subseteq (\R^P)^c\).
Let \(y^\infty\in\R^U\), 
from van Lambalgen's theorem \cite{lambalgen87}, we have \(y^\infty\notin\R^{U,y^\infty}=\R^P_{y^\infty}\subseteq \R^U=\bigcap_{y\to y^\infty}\R^{P(\cdot\mid y)}\).
\qed

To show an example in which
\(\R^P_{y^\infty}\setminus\R^{P(\cdot\mid y^\infty)}\ne\emptyset\) in   (\ref{eq-singular-case}),
we construct a consistent Bayes model with the properties (i)--(iv) listed in Theorem~\ref{counter-example} 
by modifying the examples of Bauwens  \cite{Bauwens2015} (Example 2 in \cite{BST2016}).
Note that the examples in \cite{{Bauwens2015},{BST2016}} do not  imply (iv) in Theorem~\ref{counter-example}.

\begin{figure}[ht]
\centering
\begin{picture}(200,170)(0,0)
\put(20,20){\framebox(140,140)}
\put(175,20){\(Y\)}
\put(0,160){\(X\)}
\put(20,50){\line(1,0){140}}

\put(20,70){\line(1,0){140}}

\put(20,100){\line(1,0){140}}

\multiput(143,80)(0,4){3}{\circle*{2}}

\put(0,30){\(B_1\)}
\put(0,58){\(B_2\)}
\put(0,100){\(\alpha\)}
{\small
\put(86,3){\(10^\infty\)}
\put(118,3){\(110^\infty\)}
\put(160,3){\(1^\infty\)}
}
\linethickness{1.6\unitlength}
\put(125,50){\line(0,1){20}}
\put(90,20){\line(0,1){30}}
\put(160,100){\line(0,1){60}}

\end{picture}
\caption{Construction of the counter-example.
The joint measure \(P\) concentrates on the thick lines.
\(P\) is consistent.
The conditional distribution \(P(\cdot\mid y^\infty)\) is not  continuous at  the parameter \(y^\infty=1^{\infty}\).
}\label{figC}
\end{figure}

\begin{theorem}\label{counter-example}
There is \((X\times Y,\B_2,P)\) such that \\
\textnormal{(i)} \(P\) is computable, \\
\textnormal{(ii)} \(\{1^\infty\}\in\R^{P_Y}\) and  \(P(\cdot\mid 1^\infty)\) is not computable, \\
\textnormal{(iii)} \(P(\cdot\mid y)\perp P(\cdot\mid z)\) for all \(\Delta(y)\cap\Delta(z)=\emptyset\), and \\
\textnormal{(iv)} \(\R^P_{1^\infty}\setminus \R^{P(\cdot\mid 1^\infty)}\ne\emptyset\) and 
\(\R^{P_X}\setminus \bigcup_{y^\infty\in\R^{P_Y}}\R^{P(\cdot\mid y^\infty)}\ne\emptyset\).
\end{theorem}
Here, \(1^\infty\) is the sequence consisting of 1s.
\bigskip\\
\noindent
Proof of Theorem~\ref{counter-example})
First, we prove (i).
Let \(P_X\) be the uniform distribution on \(X\).
Let \(\alpha^{\prime}=\sum_{s\in S}2^{-K(s)}\), where \(K\) is the prefix complexity.
Let  \(\alpha\in\Omega\) be the binary expansion of \(\alpha^{\prime}\) then  \(\alpha\in\R^{P_X}\) \cite{LV4th}.
Let   \(g(s):= \sum_{1\leq i\leq k}2^{-i}s_i\) where \(s=s_1\cdots s_k\) and \(s_i=0\text{ or }1\) for \(i=1,\ldots,k\).
Consider a computable sequence of strings \(a_1,a_2,\ldots\) such that  \(g(a_1)<g(a_2)<\cdots \) is increasing
 and \(\lim_i g(a_i)=\alpha^{\prime}\).

For all \(i\), set \(A_i:= \{ s\in S\mid g(s)\leq g(a_i)\}\), \(B_i:= \tilde{A}_i\setminus \tilde{A}_{i-1}\) for \(i\geq 1\), and
\(A_0=\emptyset\).
We define the measure \(P\) on \(X\times Y\) as follows.
For all \(x\in S\) and \(i\), define
\begin{align*}
&P(\lambda,0):= 0,\\
&P((B_i\cap \Delta(x))\times \{1^i0^\infty\}):= P_X(B_i\cap\Delta(x)),\\
&P(B_i^c\times \{1^i0^\infty\}):= 0, \\\
&P(((\tilde{A}_{i-1})^c\cap \Delta(x))\times \Delta(1^i)):= P_X((\tilde{A}_{i-1})^c\cap \Delta(x)), \text{and}\\
&P(\tilde{A}_{i-1}\times \Delta(1^i)):= 0,
\end{align*}
where \(1^i\) is the string consisting of \(i\) 1s.
For the construction of the measure, see Fig.~\ref{figC}.
By construction, the total measure of \(P\) is one.
Let \(C:= \bigcup_i B_i\times \Delta(1^i0)\).
For all \(x,y\in S\), we have
\begin{align*}
&P(x,y)=P((\Delta(x)\times\Delta(y))\cap C)\text{ if }y\ne 1^{|y|}, \\
&P(x,y)=P_X(\Delta(x)\cap (\tilde{A}_{|y|-1})^c)\text{ if }y= 1^{|y|}.
\end{align*}
Since the right-hand sides are computable, \(P\) is computable. 

Next, we show (ii).
Since \(P(\Omega\times \Delta(1^k))=P_X((\tilde{A}_{k-1})^c)\), we have \(P_Y(\{1^\infty\})=P(\Omega\times \Delta(1^\infty))=P_X(\cap_k (\tilde{A}_k)^c)=1-\alpha>0\) and \(\{1^\infty\}\in \R^{P_Y}\). 
We have
\(P(x\mid 1^\infty)=0\) if \(\Delta(x)\subseteq \cup_i \tilde{A}_i\) and 
 \(P(x\mid 1^\infty)=P_X(x)/(1-\alpha)\) if \(\Delta(x)\subseteq \cap_i (\tilde{A}_i)^c\).
 Since \(\alpha\) is not computable, \(P(\cdot\mid 1^\infty)\) is not computable. 
 
 By construction, we have \(P_Y(\{1^i0^\infty\})>0\) for all \(i\) and \(P_Y(\{1^\infty\})>0\).
Since \(P_Y((\bigcup_i \{1^i0^\infty\} \cup \{1^\infty\})^c)=0\),  the set \((\bigcup_i \{1^i0^\infty\} \cup \{1^\infty\})^c\) is covered by a test, and we have \(\R^{P_Y}=\bigcup_i \{1^i0^\infty\} \cup \{1^\infty\}\).
By construction, \(P(\cdot\mid y^\infty)\) is orthogonal for different \(y^\infty\in\R^{P_Y}\); we obtain   statement (iii) of the theorem. 
 
Similar to \cite{{BST2016},{Bauwens2015}}, 
 set \(U_n:= \{s\mid \exists i\ g(s)<g(a_i)+\frac{1}{n}\}\). Then \(U=\{ (s,n)\mid \exists n\ s\in U_n\}\) is r.e.~and \(P(\tilde{U}_n\mid 1^\infty)\leq 1/n\) for all \(n\).  
 \(U\) is a test that covers \(\alpha\), i.e.~\(\alpha\notin\R^{P(\cdot\mid 1^\infty)}\).

Let \(b_1\sqsubset b_2\cdots\) be an increasing sequence of prefixes  of \(\alpha\) such that \(b_n\to \alpha\) as \(n\to\infty\).
By construction, \(\forall k\exists M\forall n\geq M\ P_{Y|X}(1^k\mid b_n)=1\).
We have \(\forall k\ P_{Y|X}(1^k\mid\alpha)=1\),
 \(P_{Y|X}(\{1^\infty\}\mid\alpha)=1\), and \(\{1^\infty\}\in\R^{P_{Y|X}(\cdot\mid \alpha),\alpha}\).
Since  \(\alpha\in\R^{P_X}\),
from (\ref{genLambalA}), we have \((\alpha, 1^\infty)\in\R^P\) and \(\alpha\in\R^P_{1^\infty}\).
Since \(\alpha\notin\R^{P(\cdot\mid 1^\infty)}\), from Theorem~\ref{mutual-singular}, we have  
the first part of statement (iv) of the theorem.

Since \(\alpha\in\R^P_{1^\infty}\), \(\alpha\notin\R^{P(\cdot\mid 1^\infty)}\),  \(\R^P_{y^\infty}\) are disjoint for different \(y^\infty\in\R^{P_Y}\) (Theorem~\ref{th-consis-pos}), and 
\(\R^P_{y^\infty}\supseteq \R^{P(\cdot\mid y^\infty)}\) for all \(y^\infty\in\R^{P_Y}\) (Theorem~\ref{mutual-singular}), 
we have the latter part of statement  (iv) of the theorem. 
\qed\\

\noindent
Proof of Theorem~\ref{thMixcond}.
Part 1 follows from 
\begin{align}
\R^{P_X}=\bigcup_{y^\infty\in\R^{P_Y}}\R^P_{y^\infty} &\subseteq \bigcup_{y^\infty\in\R^{P_Y}}\bigcap_{y\to y^\infty}\R^{P(\cdot\mid y)}\label{subthMixcondA}\\
&\subseteq \bigcup_{y^\infty\in\R^{P_Y}}\R^{P_X}=\R^{P_X}, \label{subthMixcondB}
\end{align}
where (\ref{subthMixcondA}) follows from (\ref{subSet}), and (\ref{subthMixcondB}) follows from (\ref{bayes-mixtureB}).
By Lemma~\ref{lemma-A}, we have \(\R^{P_X}\supseteq\bigcup_{y^\infty\in\R^{P_Y}}\R^{P(\cdot\mid y^\infty)}\supseteq\bigcup_{y^\infty\in\R^{P_Y}}\R^{P(\cdot\mid y^\infty), y^\infty}\).\\
Part 2 follows from part 1 and Theorem~\ref{col-ifpart}.\\
Proof of part 3.
The if part follows from part 1.
The only if part follows from Theorem~\ref{th-consis-pos} (iii) and (\ref{eq-singular-case}).\\
Proof of part 4. 
Let \(P_Y(\{0^\infty\}):= 1\) and \(P(\cdot\mid 0^\infty):= U\), where \(U\) is the uniform measure on \(X\).
Then  \((X\times Y,\B_2,P)\) is computable and consistent.
We have \(P_X=U\) and \(\R^{P_X}=\R^U=\bigcup_{y^\infty\in\R^{P_Y}}\R^{P(\cdot\mid y^\infty)}=
\bigcup_{y^\infty\in\R^{P_Y}}\R^{P(\cdot\mid y^\infty),y^\infty}=
\R^{P(\cdot\mid 0^\infty)}=\R^{P(\cdot\mid 0^\infty),0^\infty}\).
Let \((X\times Y,\B_2,P^\prime)\) be the computable and consistent model defined in Theorem~\ref{counter-example}.
Then \(P_X=P^\prime_X\) and 
 \(\bigcup_{y^\infty\in\R^{P_Y}}\R^{P(\cdot\mid y^\infty)}=
\bigcup_{y^\infty\in\R^{P_Y}}\R^{P(\cdot\mid y^\infty),y^\infty}  
 \ne\bigcup_{y^\infty\in\R^\prime_Y}\R^{P^\prime (\cdot\mid y^\infty)}=
\bigcup_{y^\infty\in\R^\prime_Y}\R^{P^\prime (\cdot\mid y^\infty),y^\infty}\).
We have the proof.\\
Proof of part 5.
By definition of ML-random sequences, the sets (i), (iii), and (iv) are Borel sets.
By definition of ML-random points, \(\R^P\in\B_2\).
By Fubini theorem, \(\R^P_{y^\infty}\in\B_1\).
By Theorem~\ref{col-ifpart} and \ref{mutual-singular}, the set (iv) is the smallest one among the sets (i)--(iv),
and we have the first part of part 1.
The latter part follows from part 1.
\qed
\begin{remark}\label{remarkopen}
In Theorem~\ref{counter-example},
\(P(\cdot\mid 1^\infty)\) is not computable with oracle \(1^\infty\), 
 however 
 \(\R^{P(\cdot\mid 1^\infty)}=\R^{P(\cdot\mid 1^\infty),1^\infty}\).
The author does not know if  the statement \(\forall y^\infty\in\R^{P_Y}\ \R^{P(\cdot\mid y^\infty)}=\R^{P(\cdot\mid y^\infty),y^\infty}\) is always
 true for
the consistent Bayes models. 
\end{remark}

\section{Discussions}\label{secDiscussions}
We discuss the three topics: Bayes models and uniformly computable parametric models, 
an algorithmic solution to a classical statistical problem, and randomness in statistical models.
\subsection{Bayes models and uniformly computable parametric models}\label{sec-ML-consis}
Before we show our results, we summarize several known results for uniform tests \cite{{levinUniform},{Bienvenu2011},{gacs2005},{kjoshanssen},{vovkandvyugin93}}.
Let \(\{(\Omega,\B_1, P_{y^\infty})\mid y^\infty\in \Omega\}\) be a parameterized family of distributions 
(parametric model \(\{P_{y^\infty} \mid y^\infty\in \Omega\}\) for short).
We assume that \(y^\infty\to P_{y^\infty}(A)\) is measurable for any fixed \(A\in\B_1\) \cite{BreimanConiss64}
and uniformly computable \cite{{levinUniform},{Bienvenu2011},{gacs2005},{kjoshanssen},{vovkandvyugin93}}.
If \(P_{y^\infty}\) is uniformly computable, then \(P_{y^\infty}(\cdot)\) is defined for all \(y^\infty\) and the function 
\(y^\infty\to P_{y^\infty}(x)\) is  continuous for each fixed \(x\in S\).

A function \(t\colon \Omega\to [0,\infty]\) is called lower semicomputable if
the set \(\{x^\infty\mid r<t(x^\infty)\}\) is effectively open uniformly in rational \(r\). 
A lower semicomputable function \(t\colon \Omega\times\Omega\to [0,\infty]\) is defined in a similar manner \cite{Bienvenu2011}.
We call a function \(t_A\colon \Omega\to[0,\infty]\) a {\it blind test} w.r.t.~\(P\) and oracle \(A\) if \(t_A\) is lower semicomputable with oracle
\(A\) and \(\int t_A(x^\infty)dP\leq 1\). 
Let \(T(P,A)\) be the set of blind tests w.r.t.~\(P\) and oracle \(A\).
The set of blind  random sequences w.r.t.~\(P\) and oracle \(A\) is defined by
 \(\blind(P,A):=\{x^\infty\mid t_A(x^\infty)<\infty\text{ for all } t_A\in T(P,A)\}\).
The set of blind random sequences w.r.t.~\(P\), \(\blind(P)\) is defined similarly.
A lower semicomputable function \(t\colon \Omega\times\Omega\to [0,\infty]\) is called a {\it uniform test} w.r.t.~a parametric model
\(\{P_{y^\infty}\mid y^\infty\in\Omega\}\)
if \(\int t(x^\infty,y^\infty)dP_{y^\infty}(x^\infty)\leq 1\) for all \(y^\infty\in\Omega\).

The following proposition shows the relationship between \(\R^P\) (\(\R^{P,A}\)), blind randomness, and randomness for uniformly computable parametric models.
\begin{prop}[Bienvenu et al.~\cite{Bienvenu2011}]\label{MLblind}\hfill\\
1.~\(\blind(P)=\R^{P}\) and \(\blind(P,A)=\R^{P,A}\) 
 for all probability measure \(P\) on \((\Omega, \B)\) and oracle \(A\).
Assume that \(P\) are computable and  computable with oracle \(A\)
then  \(\R^P\) and \(\R^{P,A}\) coincide with Martin-L\"{o}f random sequences and 
those with oracle \(A\) \cite{martin-lof66}, respectively.\\
\noindent
2.~Assume that \(\{P_{y^\infty}\mid y^\infty\in\Omega\}\) is  uniformly computable.
 Then, there is a universal uniform test \(\bt\), i.e.~for each uniform test \(t\) w.r.t.~\(P_{y^\infty}\) for \(y^\infty\in\Omega\)
 there is \(c>0\) such that 
\(t\leq c\bt\), and  \(\{x^\infty\mid \bt(x^\infty,y^\infty)<\infty\}=
\blind(P_{y^\infty},y^\infty)=\R^{P_{y^\infty},y^\infty}\) 
for all \(y^\infty\in\Omega\).
\end{prop}
A Bayes model \((X\times Y,\B_2,P)\) is defined from a parametric model \(\{P_{y^\infty} \mid y^\infty\in \Omega\}\) and a prior 
\((Y,\B_1,P_Y)\) by
\begin{equation}\label{BayesParametmodel}
P(x,y):=\int_{\Delta(y)}P_{y^\infty}(x)dP_Y(y^\infty)\text{ for all }x,y\in S.
\end{equation}

\begin{theorem}[Vovk and Y'yugin\cite{vovkandvyugin93}]\label{thvv93}
Assume that a parametric model  \(\{P_{y^\infty}\mid y^\infty\in\Omega\}\) is uniformly computable and 
 \((Y,\B_1,P_Y)\) is computable. 
Let \((X\times Y,\B_2,P)\) be the Bayes model defined by (\ref{BayesParametmodel}).
Then
\[\R^P_{y^\infty}=\R^{P_{y^\infty},y^\infty}
\text{ for all }y^\infty\in\R^{P_Y}\text{ and }\R^{P_X}=\bigcup_{y^\infty\in\R^{P_Y}}\R^P_{y^\infty}.\]
\end{theorem}
\begin{definition}[Bienvenu et al.~\cite{Bienvenu2011}]
A uniformly computable parametric model \(\{P_{y^\infty}\mid y^\infty\in\Omega\}\) is called
{\it effectively orthogonal} if 
\(\R^{P_{y^\infty},y^\infty}\cap \R^{P_{z^\infty},z^\infty}=\emptyset\) for all \(y^\infty\ne z^\infty\).  
\end{definition}
\begin{theorem}
[Theorem~5.41 in Bienvenu et al.~\cite{Bienvenu2011}]\label{bienvenuetal}
Assume that a parametric model  \(\{P_{y^\infty}\mid y^\infty\in\Omega\}\) is uniformly computable and effectively orthogonal.
Then
\begin{equation}\label{uniform-disjoint-general}
\R^{P_{y^\infty},y^\infty}=\R^{P_{y^\infty}}\text{ for all }y^\infty.
\end{equation}
\end{theorem}

We state our results for uniform tests.
\begin{definition}
\(P(\cdot\mid y^\infty)\) is called continuous at \(y^\infty\in\R^{P_Y}\) if
\(\lim_n P(x\mid y^\infty(n))=P(x\mid y^\infty)\)
for all \(x\in S\) and for any sequence \(y^{\infty}(1),y^{\infty}(2),\ldots\in\R^{P_Y}\) 
such that  \(\lim_n y^{\infty}(n)\to y^\infty\in\R^{P_Y}\).
\end{definition}
\begin{theorem}\label{propBayesUniform}
Assume that a parametric model \(\{P_{y^\infty}\mid y^\infty\in\Omega\}\)
is uniformly computable and  \((Y,\B_1,P_Y)\) is computable. 
Let  \((X\times Y,\B_2,P)\) be the Bayes model defined by (\ref{BayesParametmodel}). Then, \\
1.~\(P\) is computable. For all \(y^\infty\in\R^{P_Y}\),
\(P(\cdot\mid y^\infty)=P_{y^\infty}(\cdot)\),
\(P(\cdot\mid y^\infty)\) is continuous and computable with oracle \(y^\infty\),
\(\R^{P_{y^\infty},y^\infty}=\R^{P(\cdot\mid y^\infty),y^\infty}\),
and \(\R^{P_{y^\infty}}=\R^{P(\cdot\mid y^\infty)}\).\\
2.~In addition, if \(\{P_{y^\infty}\mid y^\infty\in\Omega\}\) is effectively orthogonal then, \(P\) is consistent. 
\end{theorem}
Proof)
Since \(P_{y^\infty}\) is uniformly computable and \(P_Y\) is computable, \(P\) is computable.
From (\ref{BayesParametmodel}), for all \(x,y\in S\)
\begin{equation}\label{subniformapprox2}
P(x\mid y)=\int_{\Delta(y)}P_{y^\infty}(x)dP_Y(y^\infty)/P_Y(y).
\end{equation}
Since \(P_{y^\infty}\) is uniformly computable, 
\(y^\infty\to P_{y^\infty}(x)\) is continuous for all \(x\in S\).
From (\ref{subniformapprox2}),  for all \(x\in S\) and \(y^\infty\in\Omega\),
\[\forall \epsilon>0\ \exists y\ \forall y'\ \vert P(x\mid y')-P_{y^\infty}(x)\vert <\epsilon\text{ for }
y\sqsubseteq y'\sqsubset y^\infty.\]
By Theorem~\ref{conditional-prob}, for all \(x\in S\) and \(y^\infty\in\R^{P_Y}\)
\[\forall \epsilon>0\  \vert P(x\mid y^\infty)-P_{y^\infty}(x)\vert <\epsilon.\]
Since \(\epsilon>0\) is arbitrary, we have that \(P(\cdot\mid y^\infty)=P_{y^\infty}(\cdot)\)  for all \(y^\infty\in\R^{P_Y}\) and the
latter statements in part 1. Continuity of \(P(\cdot\mid y^\infty)\) follows from that of \(P_{y^\infty}\).\\
Proof of part 2. Assume that the parametric model is effectively orthogonal. By Theorem~\ref{thvv93}, 
\(\R^P_{y^\infty}\cap \R^P_{z^\infty}=\emptyset\) for all
\(y^\infty,z^\infty\in\R^{P_Y}\) and \(y^\infty\ne z^\infty\).
If \(y^\infty\notin\R^{P_Y}\), we have \(\R^P_{y^\infty}=\emptyset\), see \cite{lambalgen87} or \cite{takahashiIandC}.
By Theorem~\ref{th-consis-pos}, \(P\) is consistent. 
\qed

The following proposition shows that neither of the converses of Theorem~\ref{propBayesUniform}  is true. 
\begin{prop}\label{counterthBayesuniformparamodel2}
There is a computable and consistent \((X\times Y, \B_2,P)\) such that\\ (i)
\(P(\cdot\mid y^\infty)\) is computable with oracle \(y^\infty\) for all \(y^\infty\in\R^{P_Y}\) and\\ (ii)
\(0^\infty\in\R^{P_Y}\) and 
\(P(\cdot\mid y^\infty)\) is not continuous at \(0^\infty\).
\end{prop}
Proof)
Let \(r(0^n1^{\infty}):= 2^{-n}\) for \(n\geq 1\) and \(r(0^\infty):= 1\).
Let \(P_{y^\infty}(x):= r(y^\infty)^{\sum x_i}(1-r(y^\infty))^{|x|-\sum x_i}\) for all \(x=x_1\cdots x_{|x|}\in S, x_i\in\{0,1\}\)
if \(y^\infty=0^n1^\infty\) for \(n\geq 1\). Let
\(P_{y^\infty}(x):= 1\) for all \(x\sqsubset 1^\infty\) if \(y^\infty\ne 0^n1^\infty\) for  \(n\geq 1\).
Let \(P_Y(\{0^n1^\infty\}):= 2^{-(n+1)}\) and \(P_Y({0^\infty}):= 1/2\).
Then \(P_Y\) is a computable probability measure and \(\R^{P_Y}=\bigcup_{n\geq 1}\{0^n1^\infty\}\cup\{0^\infty\}\).
\(P_{y^\infty}\perp P_{z^\infty}\) if \(y^\infty,z^\infty\in\R^{P_Y}\) and \(y^\infty\ne z^\infty\).
Let \(P\) be a Bayes model defined by (\ref{BayesParametmodel}).
Then, \(P\) is computable and \(P(\cdot\mid y^\infty)=P_{y^\infty}\) for all \(y^\infty\in\R^{P_Y}\).
By Theorem~\ref{th-consis-pos}, \(P\) is consistent.
Since \(P(1\mid 0^n1^\infty)=2^{-n}\) and 
\(P(1\mid 0^\infty)=1\), \(P(\cdot\mid y^\infty)\) is not continuous at \(0^\infty\in\R^{P_Y}\).
\qed

By Theorem~\ref{propBayesUniform} and Proposition~\ref{counterthBayesuniformparamodel2}, 
the assumption of Theorem~\ref{col-ifpart} and \ref{bayes-mix} is weaker than that of Theorem~\ref{thvv93}.
 In \cite{vovkandvyugin93}, a quantitative version of Theorem~\ref{thvv93} is shown, for more details, see Section 7 in \cite{BST2016} and \cite{romachenko22}.
The equation
(\ref{singularEquality}) in Theorem~\ref{mutual-singular} is true under computable consistent Bayes models.
By Theorem~\ref{propBayesUniform} and Proposition~\ref{counterthBayesuniformparamodel2}, those models 
 constitute a larger family of models than those defined by (\ref{BayesParametmodel}) with uniformly computable and effectively orthogonal parametric models and computable priors. 
Note that the equality in (\ref{uniform-disjoint-general}) is true for all \(y^\infty\in\Omega\) while that
in  (\ref{singularEquality}) is true for all \(y^\infty\in\R^{P_Y}\).
In \cite{Bienvenu2011}, Theorem~\ref{bienvenuetal} is proved for the families of measures that are parameterized by themselves.
Kjos Hanssen  \cite{kjoshanssen}  demonstrated a different proof of   (\ref{uniform-disjoint-general}) for the  Bernoulli model.

\begin{definition}\label{defFreqConsis}
A measurable function (estimator) \(f\colon X\to Y\)  is called consistent for the parametric model
\(\{P_{y^\infty}\mid y^\infty\in\Omega\}\) if \(P_{y^\infty}(f^{-1}(y^\infty))=1\) for all \(y^\infty\in \Omega\).
A parametric model \(\{P_{y^\infty}\mid y^\infty\in\Omega\}\)
 is called consistent if there exists a consistent estimator. 
\end{definition}
In Definition 8.1 \cite{Bienvenu2011}, uniformly computable consistent models are  called orthogonal. 
We use the term consistent for the orthogonal models to distinguish pairwise orthogonality from orthogonality.

\begin{theorem}\label{paraconsistheorem}
Assume that \(\{P_{y^\infty}\mid y^\infty\in\Omega\}\) is uniformly computable. 
Let 
\[f^{-1}(y^\infty):= 
\begin{cases}
\R^{P_{y^\infty},y^\infty} \text{ if }y^\infty\ne 0^\infty,\\
\R^{P_{0^\infty},0^\infty} \cup
(\Omega\setminus \bigcup_{y^\infty\in\Omega} \R^{P_{y^\infty},y^\infty})   \text{ if }y^\infty= 0^\infty.
\end{cases}
\]
The following statements are equivalent.\\
(i) \(\{P_{y^\infty}\mid y^\infty\in\Omega\}\) is effectively orthogonal.\\
(ii) \(f\colon \Omega\to\Omega\) is well-defined and a consistent estimator for \(\{P_{y^\infty}\mid y^\infty\in\Omega\}\).
\end{theorem}
We give the proof after Definition~\ref{defDescSet}, Proposition~\ref{projectcompact}, and Lemma~\ref{borelB2}.
\begin{definition}\label{defDescSet}
Let \(A\) be a subset of \(X\times Y\) and \(B\) be a subset of \(Y\).
Set
\begin{align}
&\pi_X(A):= \{x^\infty\mid \exists y^\infty\in Y\ (x^\infty,y^\infty)\in A\}\text{ and}\label{defProjection}\\
&\pi_{X,B}(A):= \{x^\infty\mid \exists y^\infty\in B\ (x^\infty,y^\infty)\in A\}.\nonumber
\end{align}
A countable union of closed sets is called \(F_{\sigma}\) set.
\end{definition}
\begin{prop}[a special case of Exercise 2.3.24 \cite{srivastava} pp.61]\label{projectcompact}
Assume that \(C\subseteq X\times Y\) is closed.
Then \(\pi_X(C)\) is closed.
\end{prop}
Proof)
Let \(\alpha(1),\alpha(2),\ldots\) be a sequence of points in \(\pi_X(C)\).
Suppose that \(\alpha(i)\to\alpha\) as \(i\to \infty\). We show that \(\alpha\in\pi_X(C)\).
By (\ref{defProjection}), there is a sequence 
\((\alpha(1),\beta(1)),(\alpha(2),\beta(2)),\ldots\in C\). Since \(X\times Y\) is compact, \(C\) is compact and there is a subsequence\\
\((\alpha(n(1)),\beta(n(1))),(\alpha(n(2)),\beta(n(2))),\ldots\in C\) such that 
\(n(1)<n(2)<\cdots\) and 
\((\alpha(n(i)),\beta(n(i)))\to (\alpha^\prime,\beta^\prime)\in C\) as \(i\to\infty\).
Thus
\(\lim_i \alpha(i) =\lim_i \alpha(n(i))=\alpha^\prime=\alpha\in\pi_X(C)\).
\qed
\begin{lemma}\label{borelB2}
Assume that \(A\) is  \(F_{\sigma}\) in \(X\times Y\).
Then \(\pi_X(A)\) is \(F_{\sigma}\). 
Further assume that \(\pi_{X,C}(A)\cap \pi_{X,D}(A)=\emptyset\) for all \(C\cap D=\emptyset\).
Then  \(\pi_{X,C}(A)\in\B_1\) for all \(C\in\B_1\).
\end{lemma}
Proof)
Since \(A\) is \(F_\sigma\), there are closed sets \(A_n\) in \(X\times Y\) for all \(n\) and
\(A=\bigcup_n A_n\).
First we show that \(\pi_X(A)\in\B_1\).
We have
\[
\pi_X(A)=\bigcup_{y^\infty\in Y}A_{y^\infty}=\bigcup_{y^\infty\in Y}\bigcup_n (A_n)_{y^\infty}
=\bigcup_n\bigcup_{y^\infty\in Y} (A_n)_{y^\infty}=\bigcup_n\pi_X(A_n).
\]
By Proposition~\ref{projectcompact}, \(\pi_X (A_n)\) is closed and  \(\pi_X(A)\) is \(F_\sigma\).
Similarly, \\
1.~\(\pi_{X,\Delta(s)}(A)\) is \(F_\sigma\) for all \(s\in S\).\\
2.~Assume that \(\pi_{X,C_k}(A)\in\B_1\) for all \(k\).
Then \(\pi_{X,\bigcup_k C_k}(A)=\bigcup_k \pi_{X,C_k}(A)\in\B_1\).\\
3.~Assume that \(\pi_{X,C}(A)\in\B_1\). By assumption, \(\pi_{X,C^c}(A)=\pi_{X}(A)\setminus \pi_{X,C}(A)\in\B_1\).\\
Let \(\B^{\prime}:= \{C\mid \pi_{X,C}(A)\in\B_1\}\). From 1--3, 
\(\{\Delta(s)\mid s\in S\}\subseteq\B^\prime\) and
\(\B^\prime\) is a \(\sigma\)-algebra.
By definition, we have \(\B_1\subseteq\B^\prime\).
\qed

\noindent
Proof of Theorem~\ref{paraconsistheorem})
First we prove (i) \(\Rightarrow\) (ii).
Let \(\bt\) be a universal uniform test and \(A_n:= \{(x^\infty,y^\infty)\mid \bt(x^\infty,y^\infty)>n\}\) for all \(n\).
Then \(A_n\) is open,
 \(\R^{P_{y^\infty}, y^\infty}=((\cap_n A_n)^c)_{y^\infty}\) for all \(y^\infty\in\Omega\), and 
 \(A:=(\cap_n A_n)^c\) is a 
 \(F_\sigma\) set. 
Assume (i). Then 
  \(f\colon \Omega\to\Omega\)  is well-defined, i.e.~\(f^{-1}(y^\infty)\cap f^{-1}(z^\infty)=\emptyset\)
for all \(y^\infty\ne z^\infty\). 
Since the set
 \(A\) satisfies the assumptions of Lemma~\ref{borelB2},  \(f\) is measurable.
Thus \(f\) is consistent and we have (ii). The implication (ii) \(\Rightarrow\) (i) is obvious.
\qed

In \cite{BreimanConiss64}, necessary and sufficient conditions for consistency of parametric models are shown
for general parametric models.

\subsection{Algorithmic solution to a classical statistical problem}\label{secdiscuss}
In parametric models,  estimators that are consistent at all parameters are concerned, while 
in Bayes models, those that are consistent at almost all parameters are concerned.
The identification of the points at which the posterior distributions weakly converge constitutes the problem (see pp.4  Diaconis and Freedman  \cite{diaconisFreedman1986} and pp.24 Ghosh and Ramamoorthi \cite{bayesNonparametrics}). 

Let \((X\times Y,\B_2,P)\) be computable, where \(X=Y=\Omega\).
By Theorem~\ref{th-consis-pos}, we have an algorithmic solution to the  problem, i.e.~the 
Bayes model \(P\) is consistent if and only if 
\begin{equation}\label{corPosRand}
\R^{P_Y}\subseteq\{y^\infty\mid \text{the posterior distribution is consistent at }y^\infty\}.
\end{equation}
\begin{remark}
1.~Assume that the Bayes model is computable and consistent.
Then, by (\ref{corPosRand}), 
 we know that  the posterior distribution is always consistent at the ML-random parameters without explicitly 
computing the posterior distribution.
In other words, if the Bayes model is consistent and computable and the posterior distribution is not consistent at some parameters,
then these parameters are not ML-random.
Freedman~\cite{freedman63} and Schwartz~\cite{schwartz65} identified the sets of the consistent parameters
for smooth finite dimensional i.i.d.~(independent and identically distributed) models, see pp.4 \cite{diaconisFreedman1986} or the examples below.
Though (\ref{corPosRand}) identifies a subset of consistent parameters, 
Theorem~\ref{th-consis-pos} and (\ref{corPosRand}) give a new solution to the problem
since they  even hold true for any consistent \(P\) on \((\Omega^2,\B_2)\) 
by relativizing with oracle \(P\) without any further assumption.\\
\noindent
2.~It is straightforward to extend the equivalence of the statements except for (viii) in Theorem~\ref{th-consis-pos} to   joint probabilities  on complete separable metric spaces 
 such as \(\Rb^\infty\times \Rb^\infty\), see  Doob \cite{doob48} or pp.24 Remark 1.3.3 \cite{bayesNonparametrics}.
  For example, let  \((\Rb\times \Rb, \B)\) be a  measurable space where \(\B\) is the Borel \(\sigma\)-algebra.
Then, Theorem~\ref{th-consis-pos} holds for computable probabilities on \((\Rb\times \Rb, \B)\) by replacing open  intervals \(\Delta(s), s\in S\) in statements (iii) and (iv) in Theorem~\ref{th-consis-pos}
 with half open intervals  \( [a, b),\ -\infty\leq a<b\leq \infty,\ a,b\in\Qb\cup\{-\infty,\infty\}\).
Similarly, we can extend Theorem~\ref{th-consis-pos} to probability measures on \(\Rb^\infty\times \Rb^\infty\).
\end{remark}

We examine (\ref{corPosRand}) with examples of consistent Bayes models on complete separable metric spaces.

\begin{example}[Example 2 pp.17 in Schwartz~\cite{schwartz65}]\label{exampleSchwartz}
Let \(\Theta=[1,2)\) be the parameter space and the prior \(P_{\Theta}\) the uniform measure on  \(\Theta\).
Let \(X_1,X_2,\ldots\) be i.i.d.~random variables and
\(X^n:= (X_1,\ldots, X_n)\)  where \(X_1\) obeys the uniform measure on \([0,1)\) if 
\(\theta=1\) and on \([0,2/\theta)\) if \(1<\theta<2\). 
Let \(p_{\theta}(X^n)\) be the probability density function for \(\theta\in\Theta\).
Then \(p_{\theta}(X^n)=1\)  if \(\theta=1\) else \((\theta/2)^n\).
Let \(\hat{\theta}(X^n):= \argmax_{\theta} p_{\theta}(X^n)\) and \(Y_n:= \max_{1\leq i\leq n} X_i\).
Then \(\hat{\theta}=1\) if \(Y_n\leq 1\) else \(2/Y_n\).
 \(\hat{\theta}\) is consistent for all parameters.
The Bayes model is computable. 
By Theorem~\ref{th-consis-pos}, the Bayes model  is consistent, and we have (\ref{corPosRand}).
The expectation of the parameter w.r.t.~the posterior distribution (Bayes estimate) is 
\[
\beta_n(X^n)=
\begin{cases}
\frac{n+1}{n+2}\frac{2^{n+2}-1}{2^{n+1}-1}&\text{ if }Y_n\leq 1\\
\frac{\int_1^{2/Y_n}\theta^{n+1}d\theta}{\int_1^{2/Y_n}\theta^{n}d\theta}&\text{ else,}
\end{cases}
\]
and the posterior distribution is consistent except for \(\theta=1\).
Since \(P_\Theta\) is the uniform measure, \(1\notin\R^{P_\Theta}\), and  (\ref{corPosRand}) is true.
Let \(P^{\prime}_\Theta:= \alpha P_\Theta+(1-\alpha) \delta_{1}\), where \(0<\alpha<1\) and \(\delta_{1}(\{1\})=1\).
Then, the posterior distribution with prior \(P^{\prime}_\Theta\) is consistent for all \(\theta\).
Since \(P^{\prime}_\Theta(\{1\})>0\), \(1\in\R^{P^{\prime}_\Theta}\), and  (\ref{corPosRand}) is true.
\end{example}

Before we show Example~\ref{examplesupport}, we prove Proposition~\ref{propconstSup} for constructive topological spaces.

\begin{definition}[Constructive topological space \cite{{Bienvenu2011},{gacs2005},{martin-lof68}} ]\label{defconsttopolo}
Let  \((X,\T)\) be a topological space with countable base \(\Oc\), i.e.~every 
open set \(A\in\T\) is a union of elements of  \(\Oc\).
Let \(D\) be a subset of \(\Nb\) and  \(f\colon D\to\Oc\)  an onto function (naming system p.132 \cite{gacs2005}).
We call \((X,\T,\Oc,f,D)\) {\it constructive topological space} 
if \(D\) is r.e.
The countable base \(\Oc\) is called r.e.~if \((X,\T,\Oc,f,D)\) is a constructive topological space. 
\end{definition}

\begin{definition}[ML-randomness on constructive topological space]\label{MLrandomconsttopo}
Let \((X,\T,\Oc,f,D)\) be a constructive topological space and 
\((X,\B,P)\) a probability space where \(\B\) is the \(\sigma\)-algebra generated by \(\Oc\).
Then ML-test and \(\R^P\)  w.r.t.~\(P\) are defined with similar manner in Section~\ref{secDefCondML}, 
i.e.~\(U\subseteq D\times\Nb\) is called ML-test or test w.r.t.~\(P\) if \(U\) is r.e.~and
for all \(n\)  \(\tilde{U}_n\supseteq \tilde{U}_{n+1}\)
 and \(P(\tilde{U}_n)<2^{-n}\) where \(U_n:= \{x\mid (x,n)\in U\}\) and 
\(\tilde{A}:= \bigcup_{s\in A}f(s)\) for \(A\subseteq D\).
Then \(\R^P:= (\bigcup_{U\colon test}\bigcap_n \tilde{U}_n)^c\).
\end{definition}
\begin{prop}\label{propconstSup}
Assume that  \((Y,\T,\Oc,f,D)\) is a constructive topological space.
Let \((Y,\B,P)\) be a probability measure where \(\B\) is the \(\sigma\)-algebra generated by \(\Oc\).
Then
\begin{equation}\label{randSupport}
\R^{P}\subseteq\text{ the support of }P,
\end{equation}
where the support of \(P\) is the complement of the largest open null set w.r.t.~\(P\).
\end{prop}
Proof)
The complement of the support is a finite or countable union of null sets of \(\Oc\).
For each null set \(A\in\Oc\), there is \(i\) such that \(f(i)=A\) and \(P(A)=0\), i.e.~\(A\) is covered by a test.
Since a countable union of tests covers the complement of the support, we have the proposition. 
\qed

For simplicity,  we write  \(X\) for a constructive topological space \((X,\T,\Oc,f,D)\).
\begin{remark}
1.~\(\R^{P}\) in (\ref{randSupport}) is blind randomness since
we do not assume that \(P\) is computable in Definition~\ref{MLrandomconsttopo}.\\
2.~\(\{(a,b)\mid a<b, a,b\in\Qb\}\) is an r.e~base for \(\Rb\) with usual topology. 
Similarly, for a positive integer \(m\), \(\Rb^m\) and \(\Rb^\infty\) with usual topologies have r.e.~bases and are constructive topological spaces.
For other examples of r.e.~bases and constructive topological spaces, see \cite{gacs2005}.
\end{remark}

\begin{example}[Freedman~\cite{freedman63} Schwartz~\cite{schwartz65}]\label{examplesupport}
Freedman \cite{freedman63} and Schwartz \cite{schwartz65}
showed that for smooth i.i.d.~parametric models with finite-dimensional parameter space \(\Rb^m\),
posterior distributions are consistent on the support and inconsistent outside the support.
Since \(\Rb^m\) has an r.e.~base, 
by (\ref{randSupport}), we have (\ref{corPosRand}) for these models.
\end{example}

\subsection{Randomness in statistical models}\label{discussBayesDef}
We compare  randomness in Bayes models with that in
non-Bayes models.
\begin{example}[Uniformly computable parametric models]\label{finalExample2}
Assume that \(\{P_{y^\infty}\mid y^\infty\in\Omega\}\) is uniformly computable and effectively orthogonal.
Let \(g^{-1}(y^\infty):= \R^{P_{y^\infty},y^\infty}\) for all \(y^\infty\in\Omega\).
Then \(g\colon \bigcup_{y^\infty\in\Omega}\R^{P_{y^\infty},y^\infty}\to\Omega\) is well-defined and surjective.
By Lemma~\ref{borelB2}, \(g\) is measurable.
We have \(g(x^\infty)=y^\infty\) and \(x^\infty\in\R^{P_{y^\infty},y^\infty}\) for all 
\(x^\infty\in \bigcup_{y^\infty\in\Omega}\R^{P_{y^\infty},y^\infty}\).
\end{example}
In the example above, first we consider the set of random sequences \(\bigcup_{y^\infty\in\Omega}\R^{P_{y^\infty},y^\infty}\), 
the union of the sets of random sequences w.r.t.~the parameterized family of models.
Then for uniformly computable and effectively orthogonal models, 
for all \(x^\infty\in \bigcup_{y^\infty\in\Omega}\R^{P_{y^\infty},y^\infty}\),
the true parameter  \(y^\infty\) is estimated by \(g(x^\infty)\) (\(g(x^\infty)=y^\infty\))
 and \(x^\infty\) is random w.r.t.~\(P_{y^\infty}\) (\(x^\infty\in\R^{P_{y^\infty},y^\infty}\)).
  For a similar argument on collectives, see von Mises \cite{mises}.
  
In Bayes models, we assume that a sequence \(x^\infty\) is random w.r.t.~the marginal distribution on sample space.
Assumption (i') requires that \(x^\infty\) is also a member of 
the union of the sets of random sequences w.r.t.~conditional distributions given random parameters.
Then for consistent models,
for all random sequence \(x^\infty\) w.r.t.~the marginal distribution on sample space,
 the true parameter \(y^\infty\) is estimated by the posterior distributions given finite prefixes of \(x^\infty\), \(x^\infty\) is random w.r.t.~the conditional distribution given \(y^\infty\), and \(y^\infty\) is random w.r.t.~the prior.
In other words, we may say that our notion of  randomness is a Bayes version of that 
in uniformly computable parametric models.

\begin{center}
{\bf Acknowledgment}
\end{center}
The author wishes to thank the anonymous referees for their insightful comments, which significantly improved the paper and
helped in highlighting the relevant results and classical statistical problems. 
A part of this work was supported by KAKENHI (24540153).


\begin{thebibliography}{36}
\expandafter\ifx\csname natexlab\endcsname\relax\def\natexlab#1{#1}\fi
\providecommand{\url}[1]{\texttt{#1}}
\providecommand{\href}[2]{#2}
\providecommand{\path}[1]{#1}
\providecommand{\DOIprefix}{doi:}
\providecommand{\ArXivprefix}{arXiv:}
\providecommand{\URLprefix}{URL: }
\providecommand{\Pubmedprefix}{pmid:}
\providecommand{\doi}[1]{\href{http://dx.doi.org/#1}{\path{#1}}}
\providecommand{\Pubmed}[1]{\href{pmid:#1}{\path{#1}}}
\providecommand{\bibinfo}[2]{#2}
\ifx\xfnm\relax \def\xfnm[#1]{\unskip,\space#1}\fi
\bibitem[{Ackerman et~al.(2011)Ackerman, Freer and Roy}]{roy2011confpaper}
\bibinfo{author}{Ackerman, N.L.}, \bibinfo{author}{Freer, C.E.},
  \bibinfo{author}{Roy, D.M.}, \bibinfo{year}{2011}.
\newblock \bibinfo{title}{On the computability of conditional probability}, in:
  \bibinfo{booktitle}{IEEE 26th Annual Symposium on Logic in Computer Science},
  pp. \bibinfo{pages}{107--116}.
\newblock \bibinfo{note}{Arxiv:1005.3014}.
\bibitem[{Barron et~al.(1998)Barron, Rissanen and Yu}]{barron-et-al}
\bibinfo{author}{Barron, A.}, \bibinfo{author}{Rissanen, J.},
  \bibinfo{author}{Yu, B.}, \bibinfo{year}{1998}.
\newblock \bibinfo{title}{The minimum description length principle in coding
  and modeling}.
\newblock \bibinfo{journal}{IEEE Trans.~Inform.~Theory} \bibinfo{volume}{44},
  \bibinfo{pages}{2743--2760}.
\bibitem[{Barron(1985)}]{barronphd}
\bibinfo{author}{Barron, A.R.}, \bibinfo{year}{1985}.
\newblock \bibinfo{title}{Logically smooth density estimation}.
\newblock Ph.D. thesis. Stanford Univ.
\bibitem[{Bauwens(2017)}]{Bauwens2015}
\bibinfo{author}{Bauwens, B.}, \bibinfo{year}{2017}.
\newblock \bibinfo{title}{Conditional measure and the violation of van
  {L}ambalgen's theorem for {M}artin-{L}{\"{o}}f randomness}.
\newblock \bibinfo{journal}{Theory Comput.~Syst.} \bibinfo{volume}{60},
  \bibinfo{pages}{314--323}.
\newblock \bibinfo{note}{Arxiv:1103.1529}.
\bibitem[{Bauwens et~al.(2017)Bauwens, Shen and Takahashi}]{BST2016}
\bibinfo{author}{Bauwens, B.}, \bibinfo{author}{Shen, A.},
  \bibinfo{author}{Takahashi, H.}, \bibinfo{year}{2017}.
\newblock \bibinfo{title}{Conditional probabilities and van {L}ambalgen theorem
  revisited}.
\newblock \bibinfo{journal}{Theory Comput.~Syst.} \bibinfo{volume}{61},
  \bibinfo{pages}{1315--1336}.
\bibitem[{Bienvenu et~al.(2011)Bienvenu, G\'acs, Hoyrup, Rojas and
  Shen}]{Bienvenu2011}
\bibinfo{author}{Bienvenu, L.}, \bibinfo{author}{G\'acs, P.},
  \bibinfo{author}{Hoyrup, M.}, \bibinfo{author}{Rojas, C.},
  \bibinfo{author}{Shen, A.}, \bibinfo{year}{2011}.
\newblock \bibinfo{title}{Algorithmic tests and randomness with respect to a
  class of measures}.
\newblock \bibinfo{journal}{Proc.~of the Steklov Institute of Mathematics}
  \bibinfo{volume}{274}, \bibinfo{pages}{41--102}.
\newblock \bibinfo{note}{Arxiv:1103.1529v2}.
\bibitem[{Billingsley(1995)}]{pbmBill}
\bibinfo{author}{Billingsley, P.}, \bibinfo{year}{1995}.
\newblock \bibinfo{title}{{P}robability and {M}easures}.
\newblock \bibinfo{edition}{3rd} ed., \bibinfo{publisher}{Wiley}.
\bibitem[{Breiman et~al.(1964)Breiman, LeCam and Schwartz}]{BreimanConiss64}
\bibinfo{author}{Breiman, L.}, \bibinfo{author}{LeCam, L.},
  \bibinfo{author}{Schwartz, L.}, \bibinfo{year}{1964}.
\newblock \bibinfo{title}{Consistent estimates and zero-one sets}.
\newblock \bibinfo{journal}{Ann.~Math.~Statist.} \bibinfo{volume}{35},
  \bibinfo{pages}{157--161}.
\bibitem[{Diaconis and Freedman(1986)}]{diaconisFreedman1986}
\bibinfo{author}{Diaconis, P.}, \bibinfo{author}{Freedman, D.},
  \bibinfo{year}{1986}.
\newblock \bibinfo{title}{On the consistency of {B}ayes estimates}.
\newblock \bibinfo{journal}{Ann.~Statist.} \bibinfo{volume}{14},
  \bibinfo{pages}{1--26}.
\bibitem[{D\k{e}bowski(2009)}]{debowski}
\bibinfo{author}{D\k{e}bowski, {\L}.}, \bibinfo{year}{2009}.
\newblock \bibinfo{title}{Computable {B}ayesian compression for uniformly
  discretizable statistical models}, in: \bibinfo{booktitle}{ALT 2009}, pp.
  \bibinfo{pages}{53--67}.
\bibitem[{Doob(1948)}]{doob48}
\bibinfo{author}{Doob, J.L.}, \bibinfo{year}{1948}.
\newblock \bibinfo{title}{Application of the theory of martingales}, in:
  \bibinfo{booktitle}{Le Calcul des Probabilit{\'e}s et ses Applications.
  Colloq.~Intern.~du C.N.R.S.}, \bibinfo{address}{Paris}. pp.
  \bibinfo{pages}{22--28}.
\newblock \bibinfo{note}{The paper is available from {B}.~{L}ocker, {D}oob at
  {L}yon, www.jehps.net}.
\bibitem[{Freedman(1963)}]{freedman63}
\bibinfo{author}{Freedman, D.A.}, \bibinfo{year}{1963}.
\newblock \bibinfo{title}{On the asymptotic behavior of {B}ayes' estimates in
  the discrete case}.
\newblock \bibinfo{journal}{Ann.~Math.~Statist.} \bibinfo{volume}{34},
  \bibinfo{pages}{1386--1403}.
\bibitem[{G\'acs(2005)}]{gacs2005}
\bibinfo{author}{G\'acs, P.}, \bibinfo{year}{2005}.
\newblock \bibinfo{title}{Uniform test of algorithmic randomness over a general
  space}.
\newblock \bibinfo{journal}{Theoret.~Comput.~Sci.} \bibinfo{volume}{341},
  \bibinfo{pages}{91--137}.
\bibitem[{Ghosh and Ramamoorthi(2003)}]{bayesNonparametrics}
\bibinfo{author}{Ghosh, J.K.}, \bibinfo{author}{Ramamoorthi, R.V.},
  \bibinfo{year}{2003}.
\newblock \bibinfo{title}{Bayesian {N}onparametrics}.
\newblock \bibinfo{publisher}{Springer}.
\bibitem[{Hanssen(2010)}]{kjoshanssen}
\bibinfo{author}{Hanssen, B.K.}, \bibinfo{year}{2010}.
\newblock \bibinfo{title}{The probability distribution as a computational
  resource for randomness testing}.
\newblock \bibinfo{journal}{Journal of Logic and Analysis} \bibinfo{volume}{2},
  \bibinfo{pages}{1--13}.
\bibitem[{Kolmogorov(1933)}]{Kol33}
\bibinfo{author}{Kolmogorov, A.N.}, \bibinfo{year}{1933}.
\newblock \bibinfo{title}{Grundbegriffe der Wahrscheinlichkeitsrechnung}.
  volume~\bibinfo{volume}{2} of \textit{\bibinfo{series}{Eng.~Math.}}
\newblock \bibinfo{publisher}{Springer Verlag}, \bibinfo{address}{Berlin}.
\bibitem[{Kolmogorov(1963)}]{Kol63}
\bibinfo{author}{Kolmogorov, A.N.}, \bibinfo{year}{1963}.
\newblock \bibinfo{title}{On tables of random numbers}.
\newblock \bibinfo{journal}{Sankhy{\=a}} \bibinfo{volume}{25},
  \bibinfo{pages}{369—376}.
\bibitem[{Kolmogorov(1965)}]{Kol65}
\bibinfo{author}{Kolmogorov, A.N.}, \bibinfo{year}{1965}.
\newblock \bibinfo{title}{Three approaches to the quantitative definition of
  information}.
\newblock \bibinfo{journal}{Probl.~Inf.~Transm.} \bibinfo{volume}{1},
  \bibinfo{pages}{1--7}.
\bibitem[{Kolmogorov(1968)}]{Kol68}
\bibinfo{author}{Kolmogorov, A.N.}, \bibinfo{year}{1968}.
\newblock \bibinfo{title}{Logical basis for information theory and probability
  theory}.
\newblock \bibinfo{journal}{IEEE Trans. Inform. Theory} \bibinfo{volume}{14},
  \bibinfo{pages}{662--664}.
\bibitem[{van Lambalgen(1987)}]{lambalgen87}
\bibinfo{author}{van Lambalgen, M.}, \bibinfo{year}{1987}.
\newblock \bibinfo{title}{Random sequences}.
\newblock Ph.D. thesis. Universiteit van Amsterdam.
\bibitem[{Levin(1976)}]{levinUniform}
\bibinfo{author}{Levin, L.A.}, \bibinfo{year}{1976}.
\newblock \bibinfo{title}{Uniform tests of randomness}.
\newblock \bibinfo{journal}{Soviet Math.~Dokl.} \bibinfo{volume}{17},
  \bibinfo{pages}{337--340}.
\bibitem[{Li and Vit{\'a}nyi(2019)}]{LV4th}
\bibinfo{author}{Li, M.}, \bibinfo{author}{Vit{\'a}nyi, P.},
  \bibinfo{year}{2019}.
\newblock \bibinfo{title}{An introduction to {K}olmogorov complexity and Its
  applications}.
\newblock \bibinfo{edition}{Forth} ed., \bibinfo{publisher}{Springer},
  \bibinfo{address}{New York}.
\bibitem[{Martin-L{\"o}f(1966)}]{martin-lof66}
\bibinfo{author}{Martin-L{\"o}f, P.}, \bibinfo{year}{1966}.
\newblock \bibinfo{title}{The definition of random sequences}.
\newblock \bibinfo{journal}{Information and Control} \bibinfo{volume}{9},
  \bibinfo{pages}{602--609}.
\bibitem[{Martin-L{\"o}f(1968)}]{martin-lof68}
\bibinfo{author}{Martin-L{\"o}f, P.}, \bibinfo{year}{1968}.
\newblock \bibinfo{title}{Notes on constructive mathematics}.
\newblock \bibinfo{publisher}{Almqvist \& Wiksell},
  \bibinfo{address}{Stockholm}.
\bibitem[{von Mises(1981)}]{mises}
\bibinfo{author}{von Mises, R.}, \bibinfo{year}{1981}.
\newblock \bibinfo{title}{Probability, Statistics and Truth}.
\newblock \bibinfo{publisher}{Dover}.
\bibitem[{Romashchenko et~al.(2021)Romashchenko, Shen and
  Zimand}]{romachenko22}
\bibinfo{author}{Romashchenko, A.}, \bibinfo{author}{Shen, A.},
  \bibinfo{author}{Zimand, M.}, \bibinfo{year}{2021}.
\newblock \bibinfo{title}{27 open problems in {K}olmogorov complexity}.
\newblock \bibinfo{journal}{ACM SIGACT News} \bibinfo{volume}{52},
  \bibinfo{pages}{31--54}.
\bibitem[{Schwartz(1965)}]{schwartz65}
\bibinfo{author}{Schwartz, L.}, \bibinfo{year}{1965}.
\newblock \bibinfo{title}{On {B}ayes procedures}.
\newblock \bibinfo{journal}{Z.~Wahrscheinlichkeitstheorie} \bibinfo{volume}{4},
  \bibinfo{pages}{10--26}.
\bibitem[{Shen et~al.(2017)Shen, Uspensky and Vereshchagin}]{shenBook}
\bibinfo{author}{Shen, A.}, \bibinfo{author}{Uspensky, V.A.},
  \bibinfo{author}{Vereshchagin, N.K.}, \bibinfo{year}{2017}.
\newblock \bibinfo{title}{Kolmogorov complexity and algorithmic randomness}.
\newblock \bibinfo{publisher}{AMS}.
\bibitem[{Srivastava(1998)}]{srivastava}
\bibinfo{author}{Srivastava, S.M.}, \bibinfo{year}{1998}.
\newblock \bibinfo{title}{A course on {B}orel sets}.
\newblock \bibinfo{publisher}{Springer}.
\bibitem[{Takahashi(2006)}]{takahashiISIT2006}
\bibinfo{author}{Takahashi, H.}, \bibinfo{year}{2006}.
\newblock \bibinfo{title}{Bayesian approach to a definition of random sequences
  and its applications to statistical inference}, in: \bibinfo{booktitle}{2006
  IEEE International Symposium on Information Theory}, pp.
  \bibinfo{pages}{2180--2184}.
\bibitem[{Takahashi(2008)}]{takahashiIandC}
\bibinfo{author}{Takahashi, H.}, \bibinfo{year}{2008}.
\newblock \bibinfo{title}{On a definition of random sequences with respect to
  conditional probability}.
\newblock \bibinfo{journal}{Inform.~and Compt.} \bibinfo{volume}{206},
  \bibinfo{pages}{1375--1382}.
\bibitem[{Takahashi(2009)}]{takahashiRIMS2008}
\bibinfo{author}{Takahashi, H.}, \bibinfo{year}{2009}.
\newblock \bibinfo{title}{Some problems of algorithmic randomness on product
  space}, in: \bibinfo{booktitle}{The 8th Workshop on Stochastic Numerics},
  \bibinfo{publisher}{RIMS K{\^o}ky{\^u}roku, Kyoto University}. pp.
  \bibinfo{pages}{175--196}.
\bibitem[{Takahashi(2011)}]{takahashiIandC2}
\bibinfo{author}{Takahashi, H.}, \bibinfo{year}{2011}.
\newblock \bibinfo{title}{Algorithmic randomness and monotone complexity on
  product space}.
\newblock \bibinfo{journal}{Inform.~and Compt.} \bibinfo{volume}{209},
  \bibinfo{pages}{183--197}.
\bibitem[{Takahashi(2014)}]{takahashi2014}
\bibinfo{author}{Takahashi, H.}, \bibinfo{year}{2014}.
\newblock \bibinfo{title}{Generalization of van lambalgen's theorem and blind
  randomness for conditional probabilities}.
\newblock \bibinfo{note}{Arxiv:1310.0709}.
\bibitem[{Vovk and V'yugin(1993)}]{vovkandvyugin93}
\bibinfo{author}{Vovk, V.G.}, \bibinfo{author}{V'yugin, V.V.},
  \bibinfo{year}{1993}.
\newblock \bibinfo{title}{On the empirical validity of the {B}ayesian method}.
\newblock \bibinfo{journal}{J.~R.~Stat.~Soc. B} \bibinfo{volume}{55},
  \bibinfo{pages}{253--266}.
\bibitem[{Williams(1991)}]{williams91}
\bibinfo{author}{Williams, D.}, \bibinfo{year}{1991}.
\newblock \bibinfo{title}{Probability with {M}artingales}.
\newblock \bibinfo{publisher}{Cambridge University Press},
  \bibinfo{address}{Cambridge}.

\end{thebibliography}
\end{document}